\documentclass{pasj01}
\usepackage{lscape}
\usepackage{subfigure}

\Received{$\langle$reception date$\rangle$}
\Accepted{$\langle$acception date$\rangle$}
\Published{$\langle$publication date$\rangle$}

\begin{document}

\title{A systematic study of Galactic infrared bubbles along the Galactic plane with AKARI and Herschel}
\author{Misaki \textsc{Hanaoka}\altaffilmark{1}}%
\author{Hidehiro \textsc{Kaneda}\altaffilmark{1}}%
\author{Toyoaki \textsc{Suzuki}\altaffilmark{1}}%
\author{Takuma \textsc{Kokusho}\altaffilmark{1}}%
\author{Shinki \textsc{Oyabu}\altaffilmark{1}}%
\author{Daisuke \textsc{Ishihara}\altaffilmark{1}}%
\author{Mikito \textsc{Kohno}\altaffilmark{1}}%
\author{Takuya \textsc{Furuta}\altaffilmark{1}}%
\author{Takuro \textsc{Tsuchikawa}\altaffilmark{1}}%
\author{Futoshi \textsc{Saito}\altaffilmark{1}}%

\altaffiltext{1}{Graduate School of Science, Nagoya University, Furo-cho, Chikusa-ku, Nagoya 464-8602, Japan}
\email{hanaoka@u.phys.nagoya-u.ac.jp, kaneda@u.phys.nagoya-u.ac.jp}


\maketitle

\begin{abstract}
Galactic infrared (IR) bubbles, which have shell-like structures in the mid-IR wavelengths, are known to contain massive stars near their centers.
IR bubbles in inner Galactic regions ($|$l$|\leq$~65$^{\circ}$, $|$b$|\leq$~1$^{\circ}$) have so far been studied well to understand the massive star formation mechanisms.
In this study, we expand the research area to the whole Galactic plane (0$^{\circ}\leq$~l~$<$360$^{\circ}$, $|$b$|\leq$~5$^{\circ}$), using the AKARI all-sky survey data.
We limit our study on large bubbles with angular radii of $>$\timeform{1'} to reliably identify and characterize them.
For the 247 IR bubbles in total, we derived the radii and the covering fractions of the shells, based on the method developed in \citet{Hattori2016}.
We also created their spectral energy distributions, using the AKARI and Herschel photometric data, and decomposed them with a dust model, to obtain the total IR luminosity and the luminosity of each dust component, i.e., polycyclic aromatic hydrocarbons (PAHs), warm dust and cold dust. 
As a result, we find that there are systematic differences in the IR properties of the bubbles between inner and outer Galactic regions.
The total IR luminosities are lower in outer Galactic regions, while there is no systematic difference in the range of the shell radii between inner and outer Galactic regions.
More IR bubbles tend to be observed as broken bubbles rather than closed ones and the fractional luminosities of the PAH emission are significantly higher in outer Galactic regions.
We discuss the implications of these results for the massive stars and the interstellar environments associated with the Galactic IR bubbles.
\end{abstract}

\section{Introduction}
A large number of Galactic infrared (IR) bubbles, which have shell-like structures, are known to exist along the Galactic plane.  
Churchwell et al. (2006, 2007) cataloged about 600 objects which are located in inner Galactic regions ($|$l$|\leq$~$65^{\circ}$, $|$b$|\leq$~$1^{\circ}$), using the 8~$\mu$m band images of the Galactic Legacy Infrared Mid-plane Survey Extraordinaire (GLIMPSE; \cite{Benjamin2003}; \cite{Churchwell2009}) program with Spitzer.
In this catalog, the IR bubbles are classified into two categories, broken and closed bubbles, by their visual morphologies.
More recently, \citet{Simpson2012} created a catalog of 5106 IR bubbles with such information as the position, the radius and the thickness of each IR bubble.

The 8~$\mu$m band brightness is dominated by the emission from polycyclic aromatic hydrocarbons (PAHs), which are present ubiquitously in photodissociation regions (PDRs).
The shell structures of the IR bubbles are clearly seen in the PAH emission.
Within the PAH shells, ionized gases are distributed, emitting the 24~$\mu$m emission which traces hot dust.
\citet{Deharveng2010} investigated 102 IR bubbles which were cataloged by \citet{Churchwell2006}.
They showed that 86\% of the sample objects enclose H\emissiontype{II} regions using the Spitzer 8 and 24~$\mu$m and the radio-continuum data.
Thus, most of the IR bubbles possess ionizing massive stars near the centers of the PAH shells.

These IR bubbles are likely to be formed by the central massive stars.
Typical massive star formation mechanisms are ``collect and collapse'', ``globule squeezing'' and ``cloud-cloud collision (CCC)'' (e.g., \cite{Elmegreen1998}; Zinnecker \& Yorke 2007).
The former two mechanisms compress the interstellar media (ISM) at the edge of the H\emissiontype{II} regions and clumps by radiation from the pre-existing massive stars (e.g., \cite{Deharveng2010}; Dale et al. 2007). 
The CCC mechanism is triggered by collision between two molecular clouds, making dense cores on the collision surface.
\citet{Habe&Ohta1992} simulated the CCC process and found that head-on collisions between the clouds can produce massive stars.
Recently, several pieces of evidence for massive star formation likely triggered by CCC have been observed for a number of IR bubbles (e.g., \cite{Torii2015}; \cite{Baug2016}; \cite{Hattori2016}; \cite{Ohama2018}; \cite{Fukui2018}).
Massive stars are often obscured heavily and thus mid- and far-IR observations are crucial. 

In the previous study (\cite{Hattori2016}), the IR flux densities of each bubble were estimated by using six images in the 9, 18, 65, 90, 140 and 160~$\mu$m bands of the AKARI all-sky survey data.
They showed a tight correlation between the total IR luminosity and the shell radius, which followed the conventional picture of the Str$\ddot{\rm{o}}$mgren sphere.
They also obtained the central position and the shell radius of each object and established the quantitative criteria for classification of the shell morphologies.
Then, they found that large broken bubbles tend to have higher total IR luminosities, lower fractional luminosities of the PAH emission and dust heating sources located nearer to the shells.
Based on these results, \citet{Hattori2016} suggested that many of the large broken bubbles might have been formed by the CCC mechanism.

In the previous studies including ours, IR bubbles in inner Galactic regions ($|$l$|\leq$\timeform{65D}, $|$b$|\leq$\timeform{1D}) have been investigated intensively, whereas those in outer Galactic regions have not been investigated.
We expand the previous studies to the whole Galactic plane (0$^{\circ}\leq$~l~$<$~360$^{\circ}$, $|$b$|\leq$\timeform{5D}) to obtain the morphologies and the IR luminosities of the IR bubbles in outer Galactic regions as well.
We also add the far-IR and submillimeter wavelength data using the Herschel infrared Galactic plane survey (Hi-GAL; Molinari et al. 2010, 2016) to improve the estimation of the IR luminosity.
Then, we obtain the properties of the IR bubbles along the whole Galactic plane and discuss the effects of the massive stars and the interstellar environments on the properties of the IR bubbles in inner and outer Galactic regions.

\section{Observation and data analysis}
\label{method}
\begin{figure}[t]
  \begin{center}
    \includegraphics[width=1.0\linewidth,bb=10 55 570 340,clip]{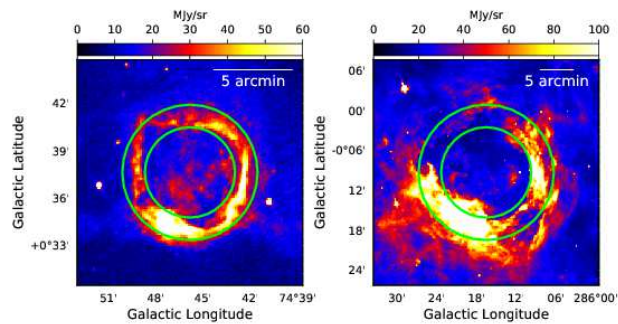}
  \end{center}
  \caption{
    Examples of the AKARI 9 $\mu$m band images of the IR bubbles newly found in this work, (left) closed and (right) broken bubbles.
    The green annular regions are defined as the shell regions which are used for morphology characterization.
    The color scales are given in units of MJy sr$^{-1}$.
  }
  \label{fig:CB_example}
\end{figure}

We searched for new IR bubbles in outer Galactic regions ($|$l$|>$\timeform{65D}, $|$b$|\leq$\timeform{5D}) and high-latitude inner Galactic regions ($|$l$|\leq$\timeform{65D}, \timeform{1D}$<|$b$|\leq$\timeform{5D}), using the AKARI all-sky survey data in the mid-IR wavelengths (central wavelengths 9~$\mu$m and 18~$\mu$m; \cite{Onaka2007}; Ishihara et al. in preparation).
In this paper, we identify IR bubbles as objects which apparently possess shell-like structures in the AKARI 9~$\mu$m band images.
Figure~\ref{fig:CB_example} shows examples of the 9~$\mu$m band images of the IR bubbles newly found in this study.
Since the resolution of the AKARI 9~$\mu$m band image is \timeform{4''.68}, which is lower than that of the Spitzer 8~$\mu$m band image (\timeform{1''.2};  \cite{Benjamin2003}), we limit our study on large bubbles with angular radii, $r$, of $>$\timeform{1'} to reliably identify and characterize them.
We also reject some objects which are too large ($r>$\timeform{20'}) or very faint for IR bubbles, such as those with the averaged shell brightnesses lower than the sky background levels.
As a result, we newly found 179 IR bubbles which consist of 141 objects in the outer Galactic regions and 38 objects in the high-latitude inner Galactic regions.
We also derived the central position and $r$ of each IR bubble, following the same procedure as in \citet{Hattori2016}.

We characterize the morphologies of 319 IR bubbles including the objects investigated in \citet{Hattori2016} by the covering fraction (CF), rather than classify it into only two types (i.e., broken or closed) as performed in the previous studies.
First, we applied smoothing for the 9~$\mu$m images with a Gaussian kernel of 3 pixels to improve the accuracy in determining the shell morphology.
Then, in order to estimate the CF, we divided the shell region (defined as an annular region from $0.8r$ to $1.2r$ as shown in figure~\ref{fig:CB_example}) into 12 sectors.
The number of the sectors for each shell is determined so that one sector has more than 30 pixels regardless of the angular sizes of the shells.
To judge whether each sector is filled or not, we adopted the condition that 40\% of the pixels in the sector should be brighter than a certain brightness threshold.
Here, we used two levels for the brightness threshold, 20\% and 30\% of the brightness averaged over the shell region, as also used in \citet{Hattori2016}, and took the average of the two results to facilitate consistency check with the result of \citet{Hattori2016}. 
In this study, we also took into account the dependence of the CF on the phase of the sector division.
We shifted the sector-dividing positions by half a sector and re-estimated the CF.
Finally, we again took the average of the two CFs before and after changing the dividing positions.

To estimate the IR fluxes, we used the AKARI mid-IR (9 and 18~$\mu$m) and far-IR (65, 90, 140 and 160~$\mu$m; \cite{Kawada2007}; \cite{Doi2015}) all-sky survey data as well as Herschel Hi-GAL data (70, 160, 250, 350 and 500~$\mu$m).
For the objects in the high-latitude regions (\timeform{1D}$<|$b$|\leq$\timeform{5D}), we used only the AKARI data since Herschel Hi-GAL data are not available in those regions.
The photometry aperture and the background region are defined by the same procedure as in \citet{Hattori2016}, which are a circular region of $<2r$ and an annular region at from $2r$ to $4r$, respectively.
Since the target sizes are not large enough to neglect the aperture corrections in the AKARI far-IR images, we estimated the aperture correction factors for the IR bubbles with various radii, based on the encircled energies of the point spread functions created from point-like galaxies in \citet{Kokusho2017}.
The aperture correction factors thus obtained are shown in table \ref{table:correction} and we applied them to the AKARI far-IR fluxes.
On the other hand, the target sizes are large enough for AKARI in the mid-IR and Herschel, and therefore aperture corrections for those images are negligible ($<1$\%) and not applied.
Here, we considered random and systematic errors as flux uncertainties; we calculated the random errors from the background fluctuation, while we adopted the systematic errors of 10\% for the AKARI mid-IR and Herschel data (Ishihara et al. in preparation; \cite{Molinari2016}) and 15\% for the AKARI far-IR data (\cite{Takita2015}).

\begin{table}
  \caption{Correction factors of the AKARI far-IR bands for the aperture photometry of the IR bubbles with various radii.}
  \label{table:correction}
  \begin{center}
    \begin{tabular}{cccc}
      \hline
      IR bubble radius & 65 $\mu$m & 90 $\mu$m & 140 and 160 $\mu$m \\\hline
      \timeform{1'}$\leq R<$\timeform{1'.25} & 1.30 & 1.44 & 1.23 \\
      \timeform{1'.25}$\leq R<$\timeform{1'.5} & 1.19 & 1.27 & 1.12 \\
      \timeform{1'.5}$\leq R<$\timeform{1'.75} & 1.14 & 1.18 & 1.07 \\
      \timeform{1'.75}$\leq R<$\timeform{2'} & 1.09 & 1.11 & 1.04 \\
      \timeform{2'}$\leq R<$\timeform{2'.25} & 1.06 & 1.07 & 1.02 \\
      \timeform{2'.25}$\leq R<$\timeform{2'.5} & 1.03 & 1.03 & 1.01 \\\hline
    \end{tabular}
  \end{center}
\end{table}

\begin{figure}[t]
  \begin{center}
    \includegraphics[width=0.9\linewidth,bb=22 17 420 320,clip]{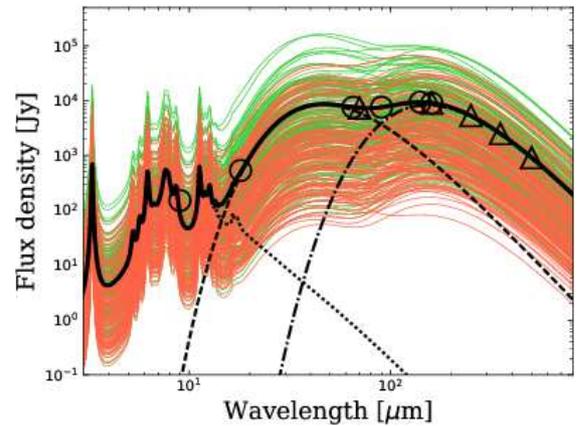}
  \end{center}
  \caption{
    Result of the SED fitting.
    The black circles and triangles are the AKARI and Herschel data points, respectively, which are averaged for all IR bubbles.
    The black solid line shows the best-fit model, while the dotted, dashed and dash-dotted lines show PAH, warm dust and cold dust components, respectively.
    The green and red solid lines correspond to the best-fit spectral models of the other IR bubbles in the inner and outer Galactic regions, respectively.
  }
  \label{fig:SED}
\end{figure}

\begin{figure}
  \begin{center}
    \includegraphics[width=1.0\linewidth,bb=10 5 410 325,clip]{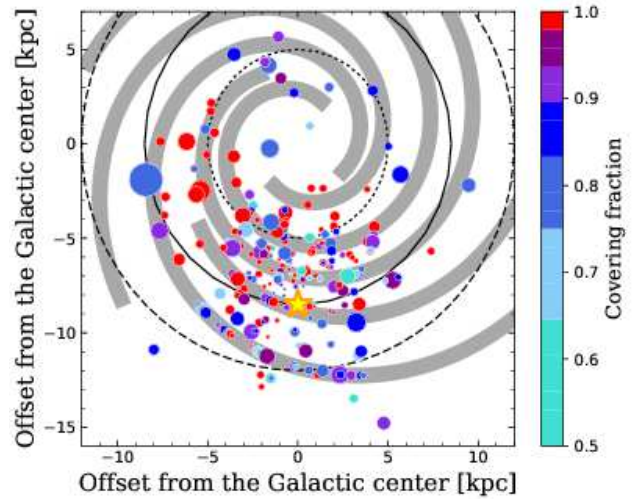}
  \end{center}
  \caption{
    Distribution of the IR bubbles plotted on the pattern of the Galactic spiral arms modeled by \citet{Nakanishi2016}.
    The colors correspond to the CF.
    The size of the symbol is proportional to the absolute shell radius of each IR bubble from 0.3 pc to 50 pc.
    The dotted, solid and dashed lines correspond to the distances of 5, 8.5 and 12 kpc from the Galactic center, respectively.
    The star symbol indicates the position of the Sun.
  }
  \label{fig:distance_innerbubble}
\end{figure}

We created the spectral energy distributions (SEDs) of the IR bubbles from the photometry fluxes, and fitted the SEDs with a dust model which includes PAHs, warm dust and cold dust components.
We adopted the PAH model by \citet{Draine&Li2007} and assumed the dust components as modified blackbody.
The emissivity power-law indices of the modified blackbody dust components are assumed to be 2 and the dust temperatures are allowed to vary for both warm and cold dust components (Anderson et al. 2012).
However, when we fit the SEDs of the IR bubbles in the high-latitude inner Galactic regions (\timeform{1D}$<|$b$|\leq$\timeform{5D}) where we have no Herschel data, we fixed the dust temperature of the cold dust at 18~K, which is a typical value of the IR bubbles analyzed in this study, due to the lack of the data points to be fitted.
Figure~\ref{fig:SED} shows the results of the SED fitting.
Based on the reduced $\chi^2$ values, about 90\% of our samples are accepted with a 90\% confidence level.
From this figure, we can recognize that the fluxes of the IR bubbles in inner Galactic regions are systematically higher than those in outer Galactic regions.

\begin{figure}[t]
  \begin{center}
    \includegraphics[width=1.0\linewidth,bb=40 20 400 495,clip]{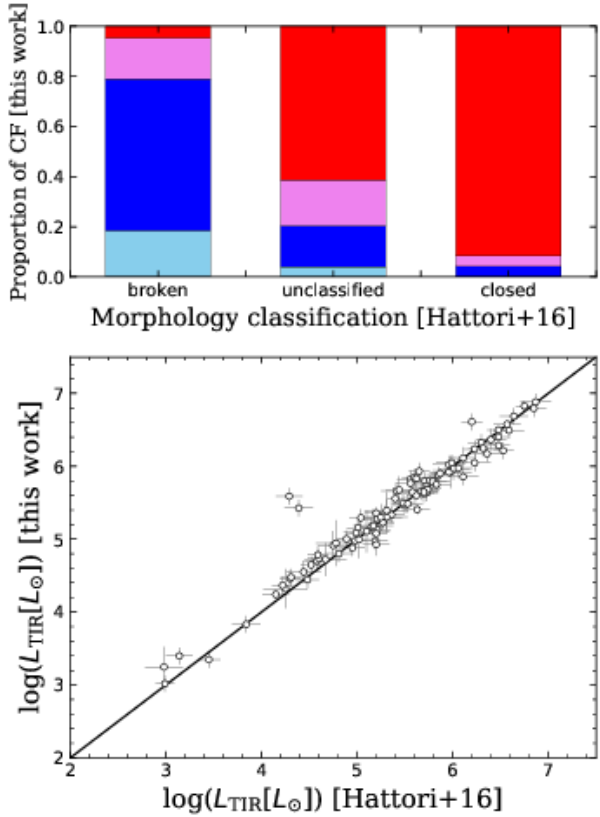}
  \end{center}
  \caption{
    Comparison between the results of this work and \citet{Hattori2016}.
    The upper panel shows the proportion of the CF.
    The red, violet, blue and cyan areas correspond to CF=$1.0$, $0.9\leq$CF$<1.0$, $0.75\leq$CF$<0.9$ and $0.5\leq$CF$<0.75$, respectively.
    The lower panel shows comparison of $L_{\rm{TIR}}$.
    The black line corresponds to $y=x$. 
    }
  \label{fig:consistency}
\end{figure}

Among the IR bubbles newly found in this study, we cross-identified 60 IR bubbles in the catalog of H\emissiontype{II} regions in Milky Way (Hou \& Han 2014) to estimate their distances.
Figure~\ref{fig:distance_innerbubble} shows the distribution of the IR bubbles with the known distance on the Galactic disk, from which we confirm that most of them are located on or near the Galactic spiral arms (Nakanishi \& Sofue 2016).
Hence, we assume that, for the IR bubbles with their distances unknown, 32 IR bubbles are located on the Perseus arm in outer Galactic regions (\timeform{90D}$<$l$<$\timeform{225D}) and 30 IR bubbles are located on the Orion and Sagittarius arms in inner Galactic regions at $|$b$|>$\timeform{2D}, toward which there is only a single arm along the line of sight.
For the 259 bubbles whose distances were thus determined, we obtained the luminosities of the SED components, $L_{\rm{PAH}}$, $L_{\rm{warm}}$, $L_{\rm{cold}}$ and $L_{\rm{TIR}}$ ($=L_{\rm{PAH}}+L_{\rm{warm}}+L_{\rm{cold}}$) as well as the absolute shell radii, $R$.
In figure~\ref{fig:consistency}, we check the consistency of our result with the result of \citet{Hattori2016}, regarding the classification of the bubble morphology and estimation of $L_{\rm{TIR}}$.
From the figure, we confirm that there is a global consistency between the two results.
For a few IR bubbles which show considerably higher $L_{\rm{TIR}}$ than those in \citet{Hattori2016}, their distances were ambiguous (\cite{Deharveng2010}; Watson et al. 2010; Kuchar \& Bania 1994), and changed from \citet{Hattori2016} to those estimated in the catalog of H\emissiontype{II} regions (Hou \& Han 2014).

The properties of all the IR bubbles detected in this study, including those in the previous studies with $r>$\timeform{1'}, are summarized in Appendix.
Note that the central positions of the IR bubbles are not necessarily close to the true positions of massive stars.
For example, massive stars are expected to lie outside bubbles, when the bubbles are part of a large irregular cavity created by the massive star.
Such cases are, ``ES6'', ``E58'', ``E97'' and ``E109'' (Deharveng et al. 2012; Dubner et al. 1992; Bassino et al. 1982).
Also note that objects such as planetary nebulae (PNe), luminous blue variables (LBVs) and supernova remnants (SNRs) can be picked up as IR bubbles mistakenly.
We have confirmed that ``E12'' and ``E43'' belong to LBVs and SNRs, respectively (Kraemer et al. 2010; Acero et al. 2016), which are removed from our sample, while PNe are not included in our sample based on the PNe catalogs (Parker et al. 2006; Miszalski et al. 2008).
Furthermore, when H\emissiontype{II} regions have bipolar morphology, they can also be mistaken as two independent bubbles (e.g., Deharveng et al. 2015; Samal et al. 2018).
We have confirmed that 10 bubbles belong to 5 bipolar H\emissiontype{II} regions (``S18'' and ``S20''; Samal et al. 2018, ``S97''; Deharveng et al. 2015, ``S109'', ``S110'' and ``S111''; Dalgleish et al. 2018, ``CN107'' and ``CN109''; Bally et al. 1983; Dewangan et al. 2016, ``EN13'' and ``EN14''; Mallick et al. 2013), which are also removed from our sample.
As a result, 247 IR bubbles remain in our final sample.
We have confirmed that 241 out of the 247 IR bubbles show the significant presence of diffuse emission at 18~$\mu$m within the shell boundaries.
\citet{Deharveng2010}, in their study of Galactic bubbles, find that extended 24 $\mu$m emission often lies close to the exciting star or cluster of the bubbles.
Therefore, almost all the IR bubbles in our sample are likely to be indeed associated with massive stars.

\section{Result}
\begin{figure}
  \begin{center}
    \includegraphics[width=1.0\linewidth,bb=30 30 400 330,clip]{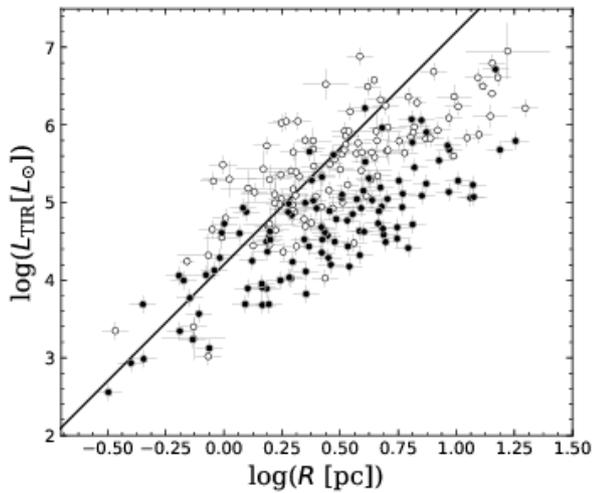}
  \end{center}
  \caption{
    Relation between $L_{\rm{TIR}}$ and the shell radius of each bubble.
    The filled symbols correspond to the IR bubbles newly found in this study, while the open symbols correspond to those investigated in \citet{Hattori2016}.
    The black line corresponds to the best-fit result of \citet{Hattori2016} with $L_{\rm{TIR}}=aR^3$.
  }
  \label{fig:Stromgren_inout}
\end{figure} 

Figure~\ref{fig:Stromgren_inout} shows the relation between $L_{\rm{TIR}}$ and $R$.
From the figure, we confirm that the result of our analysis for the IR bubbles treated in the previous study (open symbols) is consistent with the best-fit result of \citet{Hattori2016} with $L_{\rm{TIR}}=aR^3$.
Here, \citet{Hattori2016} assumed that $L_{\rm{TIR}}$ is proportional to $Q$, the total number rate of ionizing photons from central stars, and $Q$ is described as the following equation (Str$\ddot{\rm{o}}$mgren 1939):
\begin{equation}
  Q=\frac{4\pi}{3} R_{\rm{S}}^3 n_e n_p \alpha_{\rm{B}}(T_{e}),
  \label{eq:stromgren}
\end{equation} 
where $R_{\rm{S}}$, $n_e$, $n_p$, $T_{e}$ and $\alpha_{\rm{B}}$ are the Str$\ddot{\rm{o}}$mgren sphere radius, electron density, proton density, electron temperature and ``case-B'' recombination coefficient (Osterbrock 1989), respectively.
Hence, the IR bubbles in the previous study follow the conventional picture of the Str$\ddot{\rm{o}}$mgren sphere.
However, we find a systematic difference between the IR bubbles newly found in this study (filled symbols) and those investigated in the previous study; the former shows $L_{\rm{TIR}}$ significantly lower than the latter, while both show no systematic difference in the range of $R$.
Moreover, the former IR bubbles do not apparently follow the picture of the Str$\ddot{\rm{o}}$mgren sphere (i.e., $L_{\rm{TIR}} \propto R^3$).

\begin{figure}
  \begin{center}
    \includegraphics[width=1.0\linewidth,bb=25 37 440 320,clip]{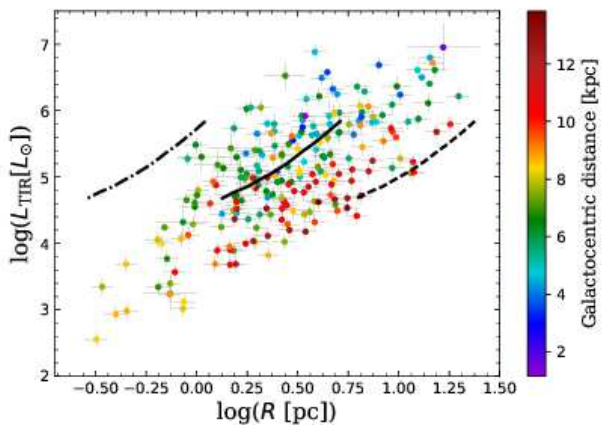}
  \end{center}
  \caption{
    Same as figure~\ref{fig:Stromgren_inout}, but the data points are color-coded according to the Galactocentric distance.
    The black dash-dotted, solid and dashed lines show the relations calculated for O3V$-$O9V stars (Martins et al. 2005) in H\emissiontype{II} regions with the electron densities of 1000, 100 and 10 cm$^{-3}$, respectively, for the temperature of $10^4$ K.
  }
  \label{fig:Stromgren_GCdistance_color}
\end{figure}      

\begin{figure}
  \begin{center}
    \includegraphics[width=1.0\linewidth,bb=10 35 440 710,clip]{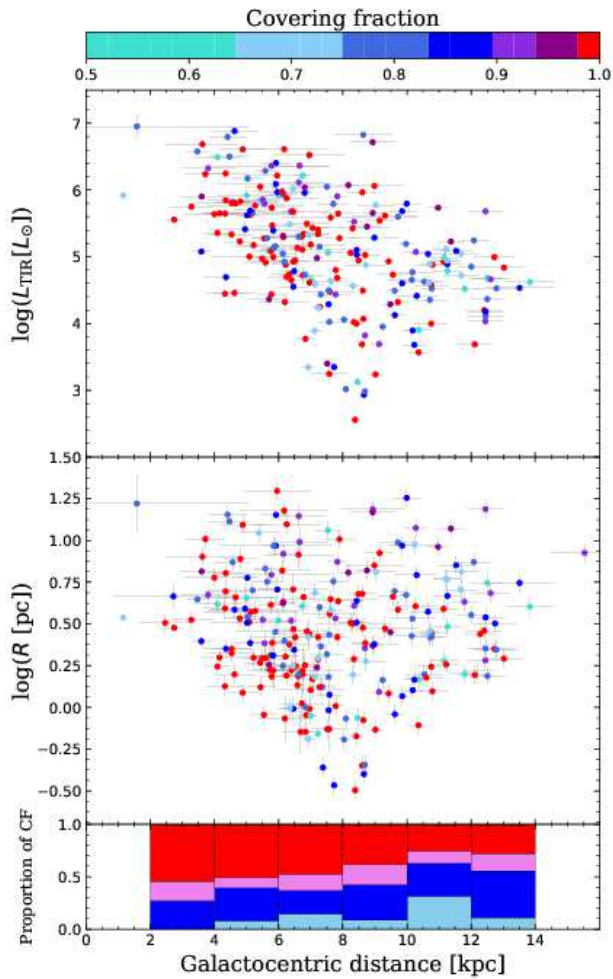}
  \end{center}
  \caption{
    Upper and middle panels show the distributions of $L_{\rm{TIR}}$ and $R$, respectively, as a function of the Galactocentric distance, color-coded according to the CF.
    The lower panel shows the proportion of the CF every $2$ kpc bin.
    The red, violet, blue and cyan areas correspond to CF=$1.0$, $0.9\leq$CF$<1.0$, $0.75\leq$CF$<0.9$ and $0.5\leq$CF$<0.75$, respectively.
  }
  \label{fig:CF_GCdistance}
\end{figure}

Most of the IR bubbles newly found in this study are located in outer Galactic regions, while all the previous samples are in inner Galactic regions.
Therefore, the systematic difference in figure~\ref{fig:Stromgren_inout} is likely to be caused by the dependence of $L_{\rm{TIR}}$ on the Galactocentric distance.
In figure~\ref{fig:Stromgren_GCdistance_color}, we color-coded the data points of the $L_{\rm{TIR}}$-$R$ relation, based on the Galactocentric distance.
The figure indeed shows the $L_{\rm{TIR}}$-$R$ relation depends on the Galactocentric distance and their correlation becomes tighter for a limited range of the distance.
Furthermore, the upper panel of figure~\ref{fig:CF_GCdistance} shows that $L_{\rm{TIR}}$ monotonically decreases with the Galactocentric distance, except for the local minimum in the solar neighborhood ($\sim$8~kpc) where relatively faint IR bubbles are detected because of their proximity.
Considering that $L_{\rm{TIR}}$ is roughly approximated by the bolometric luminosities of the massive stars associated with the IR bubbles, this result indicates that the central stars tend to be of earlier spectral types in inner Galactic regions.
\citet{Martins2005} showed that log($L_{\rm{TIR}}/$L$_{\odot}$) of a star of spectral type O9V$-$O3V ranges from 4.77 to 5.84.
The $L_{\rm{TIR}}$ values in figure~\ref{fig:CF_GCdistance} suggest that almost all the IR bubbles in inner Galactic regions ($\lesssim 7$~kpc) contain an O-type star and some of them may have tens of O-type stars, whereas a significant fraction of the IR bubbles in outer Galactic regions ($\gtrsim 9$~kpc) are not IR luminous enough to have a single O-type star.
Indeed, the distribution of the star forming rate is known to steadily decrease outward from a peak at $\sim$5~kpc from the Galactic center (e.g., Guesten \& Mezger 1982; \cite{Misiriotis2006}; Kennicutt \& Evans 2012).

The middle panel in figure~\ref{fig:CF_GCdistance} shows that $R$ does not clearly depend on the Galactocentric distance, except for the local minimum in the solar neighborhood where relatively small IR bubbles satisfy our size limit, $r>$\timeform{1'}.
Although $L_{\rm{TIR}}$ decreases monotonically, $R$ does not with the Galactocentric distance, which is consistent with the result in figure~\ref{fig:Stromgren_GCdistance_color}.
Moreover, the lower panel in figure~\ref{fig:CF_GCdistance} indicates that the proportion of the IR bubbles with relatively low CFs increases with the Galactocentric distance, i.e., the IR bubbles tend to be observed as broken bubbles in outer Galactic regions.

\begin{figure}
  \begin{center}
    \includegraphics[width=0.95\linewidth,bb= 25 40 455 330,clip]{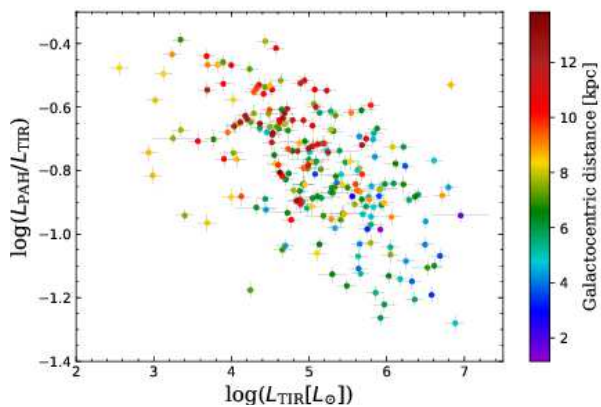}
  \end{center}
  \caption{
    Relation between $L_{\rm{PAH}}/L_{\rm{TIR}}$ and $L_{\rm{TIR}}$.
    The data points are color-coded according to the Galactocentric distance.
    }
  \label{fig:result_ratio}
\end{figure}

Figure~\ref{fig:result_ratio} shows the fractional PAH luminosities ($L_{\rm{PAH}}/L_{\rm{TIR}}$) of the IR bubbles plotted against $L_{\rm{TIR}}$, color-coded according to the Galactocentric distance.
The figure clearly exhibits a negative correlation between $L_{\rm{PAH}}/L_{\rm{TIR}}$ and $L_{\rm{TIR}}$.
As suggested by \citet{Hattori2016}, this trend can be interpreted in such a way that intense UV fluxes from central stars increase $L_{\rm{TIR}}$, which accelerate photodissociation of PAHs, thus lowering $L_{\rm{PAH}}/L_{\rm{TIR}}$.
Our new finding is that the $L_{\rm{PAH}}/L_{\rm{TIR}}$ values of the IR bubbles in outer Galactic regions are systematically higher than those in inner Galactic regions; the relation of the IR bubbles in outer Galactic regions appears to follow a trend similar to that in inner Galactic regions.
Moreover, a significant fraction of the IR bubbles in outer Galactic regions shows $L_{\rm{PAH}}/L_{\rm{TIR}}$ as high as 30$-$40\%, which is unusually high as compared to 10$-$20\% typically observed for star-forming galaxies (e.g., \cite{Smith2007}; \cite{Stierwalt2014}) as well as star-forming regions and diffuse interstellar regions in our Galaxy (e.g., \cite{Onaka1996}; \cite{Arendt1998}; \cite{Draine2011}; \cite{Kaneda2013}).
On the other hand, the ratio of the sum of $L_{\rm{PAH}}$ to the sum of $L_{\rm{TIR}}$ for all the IR bubbles along the whole Galactic plane (i.e., $\Sigma L_{\rm{PAH}}/\Sigma L_{\rm{TIR}}$) is 14\% as a whole, which is usual as typically observed values.

\section{Discussion}
\subsection{Dependence of the IR bubble properties on the Galactocentric distance}
\label{discussion:distance}
We discuss the $L_{\rm{TIR}}$-$R$ relation of the IR bubbles which shows a systematic difference between inner and outer Galactic regions.
In order to interpret the result in light of the Str$\ddot{\rm{o}}$mgren sphere picture, in figure~\ref{fig:Stromgren_GCdistance_color} we overplot the relation between the bolometric luminosity and the radius expected for a set of O3V$-$O9V stars with different electron densities (Martins et al. 2005).
We find that the $L_{\rm{TIR}}$-$R$ relation at $\sim$3 kpc corresponds to the density of $\sim$200 cm$^{-3}$ while that at $\sim$11 kpc corresponds to $\sim$30~cm$^{-3}$, and thus the required density is likely to decrease monotonically with the Galactocentric distance.
Those trends suggest that the shell may be expanded more easily in outer Galactic regions, which can explain the result that the observed range of $R$ is similar between inner and outer Galactic regions even though $L_{\rm{TIR}}$ is systematically lower in outer Galactic regions.
Considering the pressure balance between the inside and the outside of the shells, this may be related to the decline in the interstellar energy density, which is dominated by the cosmic ray and magnetic energy in the local ISM (Mathis et al. 1983; Webber \& Yushak 1983; \cite{Arendt1998}; Heiles \& Crutcher 2005; \cite{Draine2011}).
Indeed, the cosmic ray sources, which contain SNRs, pulsars and OB stars, and magnetic field strength decrease roughly by a factor of 2 from 3~kpc to 10~kpc, which is consistent with the change of the electron density and the gas pressure assuming a constant gas temperature (e.g., Case \& Bhattacharya 1998; \cite{Bronfman2000}; \cite{Han2006}; \cite{Lorimer2006}; \cite{Ackermann2012}).
This can also explain the trend of the CF decreasing with the Galactocentric distance (figure~\ref{fig:CF_GCdistance}), since the IR bubbles are expected to break more easily in the ISM of lower gas densities.

\subsection{Fractional PAH luminosities of the IR bubbles}
\label{discussion:ratio}
\begin{figure}
  \begin{center}
    \includegraphics[width=1\linewidth,bb=20 30 440 320,clip]{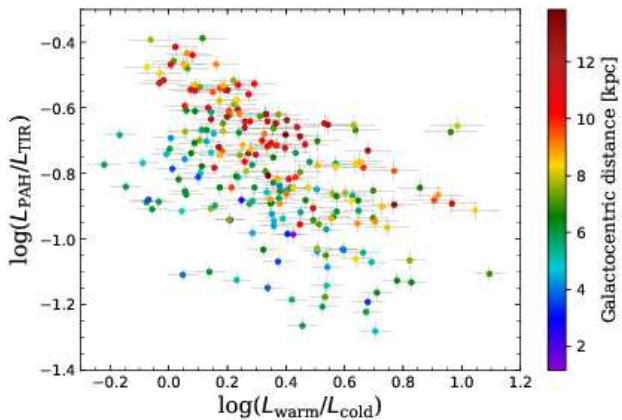}
  \end{center}
  \caption{
    Relation between $L_{\rm{PAH}}/L_{\rm{TIR}}$ and $L_{\rm{warm}}/L_{\rm{cold}}$.
    The data points are color-coded according to the Galactocentric distance.
  }
  \label{fig:PAHratio_no_extinction}
\end{figure}

We discuss the variation of the fractional PAH luminosities of the IR bubbles and its implications for the interstellar environments.
Figure~\ref{fig:result_ratio} shows a global trend of $L_{\rm{PAH}}/L_{\rm{TIR}}$ decreasing with $L_{\rm{TIR}}$.
However, as far as the IR bubbles in outer ($\gtrsim 8$ kpc) Galactic regions are concerned, dependence of $L_{\rm{PAH}}/L_{\rm{TIR}}$ on $L_{\rm{TIR}}$ is not significant.
In figure~\ref{fig:PAHratio_no_extinction}, we plot $L_{\rm{PAH}}/L_{\rm{TIR}}$ against $L_{\rm{warm}}/L_{\rm{cold}}$ instead of $L_{\rm{TIR}}$.
The figure shows the dependence of $L_{\rm{PAH}}/L_{\rm{TIR}}$ on $L_{\rm{warm}}/L_{\rm{cold}}$ more clearly than on $L_{\rm{TIR}}$ for both inner and outer Galactic regions.
This is probably because $L_{\rm{warm}}/L_{\rm{cold}}$ (i.e., dust color temperature) indicates the strength of the UV radiation exposed to the PAHs more directly than $L_{\rm{TIR}}$ (e.g., \cite{Tielens2008}).
Therefore, the negative correlation between $L_{\rm{PAH}}/L_{\rm{TIR}}$ and $L_{\rm{warm}}/L_{\rm{cold}}$ in figure~\ref{fig:PAHratio_no_extinction} is likely to be caused by photodestruction of PAHs.

\begin{figure}
  \begin{center}
    \includegraphics[width=1\linewidth,bb=20 30 440 320,clip]{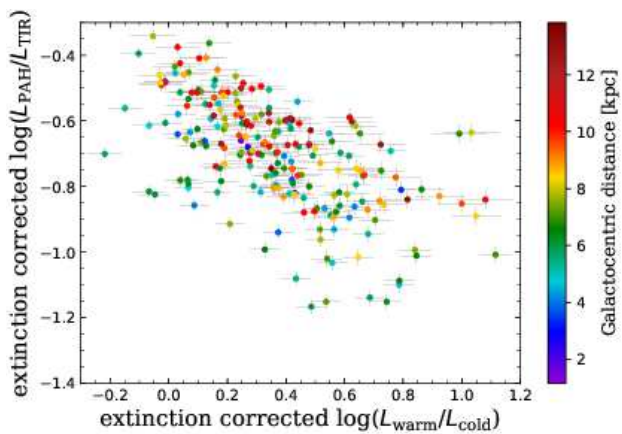}
  \end{center}
  \caption{
    Same as figure \ref{fig:PAHratio_no_extinction}, but modified considering the effect of the interstellar extinction for $L_{\rm{PAH}}/L_{\rm{TIR}}$ and $L_{\rm{warm}}/L_{\rm{cold}}$. 
  }
  \label{fig:PAHratio_distance}
\end{figure}

More importantly, figure~\ref{fig:PAHratio_no_extinction} clearly shows a systematic difference in $L_{\rm{PAH}}/L_{\rm{TIR}}$ between inner and outer Galactic regions.
Under the assumption that the relative abundance of PAHs to total dust is constant along the Galactic plane, there are mainly two causes to systematically change $L_{\rm{PAH}}/L_{\rm{TIR}}$; the cold dust component in outer Galactic regions may be too cold to fully contribute to $L_{\rm{TIR}}$, or the interstellar extinction may lower the observed $L_{\rm{PAH}}$ more in inner Galactic regions.
The former possibility calls for denser cold gas surrounding the IR bubbles in outer Galactic regions, which is rather unlikely considering that the IR bubbles tend to be more easily expanded and observed as broken bubbles in outer Galactic regions.
We evaluated the effects of the interstellar extinction on $L_{\rm{PAH}}$ (and also $L_{\rm{warm}}$) as follows: first, we derive the optical depth, $\tau$, to each IR bubble.
$\tau$ is described as:
\begin{equation}
\tau = C_{\rm{ext}}(\lambda) n_{\rm{H}} D,
\end{equation}
where $C_{\rm{ext}}(\lambda)$, $n_{\rm{H}}$ and $D$ are the dust extinction cross section as a function of wavelength, the hydrogen density and the distance to each IR bubble, respectively.
We adopted $C_{\rm{ext}}(\lambda)$ given in \citet{Draine2003a} and calculated the effective values of $C_{\rm{ext}}(\lambda)$ for the AKARI 9~$\mu$m and 18~$\mu$m bands to be $1.77\times10^{-23}$~cm$^{2}$ and $1.04\times10^{-23}$~cm$^{2}$, respectively, considering their band response curves.
We took the sum of the hydrogen atomic and molecular gas densities (i.e., $n_{\rm{H}}=n_{\rm{H\emissiontype{I}}}+2n_{\rm{H}_2}$) in the calculation where the $n_{\rm{H\emissiontype{I}}}$ distribution on the Galactic plane is taken from \citet{Wolfire2003} while the $n_{\rm{H}_2}$ distribution from that estimated by \citet{Nakanishi&Sofue2006} with the ${}^{12}$CO ($J=1-0$) survey data.
We then modified the 9~$\mu$m and 18~$\mu$m fluxes with $\tau$, and fitted the SEDs again to re-estimate $L_{\rm{PAH}}$ and $L_{\rm{warm}}$.
Figure~\ref{fig:PAHratio_distance} shows the result thus modified for the interstellar extinction.
This result still shows the systematic difference persistently, although the difference between inner and outer Galactic regions is reduced.
Therefore, the interstellar extinction is not a dominant factor to cause the systematic difference in $L_{\rm{PAH}}/L_{\rm{TIR}}$ between inner and outer Galactic regions.

It is likely that the relative abundance of PAHs to the total dust is not constant along the Galactic plane.
Mass losses from asymptotic giant branch (AGB) stars are believed to make a significant contribution to the formation of dust in the ISM (e.g., \cite{Matsuura2009}).
Among them, carbon-rich (C-rich) AGB stars are considered as suppliers of carbonaceous dust including PAHs, while oxygen-rich (O-rich) AGB stars as suppliers of silicate dust (\cite{Latter1991}; \cite{Tielens2008}).
\citet{Ishihara2011} investigated the distributions of C-rich and O-rich AGB stars in our Galaxy and obtained that the C-rich AGB stars are uniformly distributed within the Galactic disk, while the O-rich AGB stars are concentrated toward the Galactic center.
Therefore, the relative contribution of the C-rich stars to the formation of dust in the ISM is higher in outer Galactic regions, which can explain the systematically higher $L_{\rm{PAH}}/L_{\rm{TIR}}$ for IR bubbles in outer Galactic regions.
Figure~\ref{fig:PAHratio_distance} also suggests that the environments in the solar neighborhood belong to those relatively rich in PAHs in outer Galactic regions.

\subsection{Properties of the IR bubbles in light of CCC} 
In the previous study, \citet{Hattori2016} suggested that many of the large broken bubbles might have been formed by the CCC mechanism, based on the observational facts that large broken bubbles tend to have higher $L_{\rm{TIR}}$ and lower $L_{\rm{PAH}}/L_{\rm{TIR}}$.
Here we also discuss this possibility, investigating the dependence of $L_{\rm{TIR}}$ and $L_{\rm{PAH}}/L_{\rm{TIR}}$ on the CF.
To remove the dependence on the Galactocentric distance as much as possible, we treat the IR bubbles in inner ($\leq 7$~kpc) and outer ($> 7$~kpc) Galactic regions, separately.

\begin{figure*}
  \begin{tabular}{cc}
    \begin{minipage}{0.45\textwidth}
      \begin{center}
        \includegraphics[width=1.05\linewidth,bb=25 40 420 555,clip]{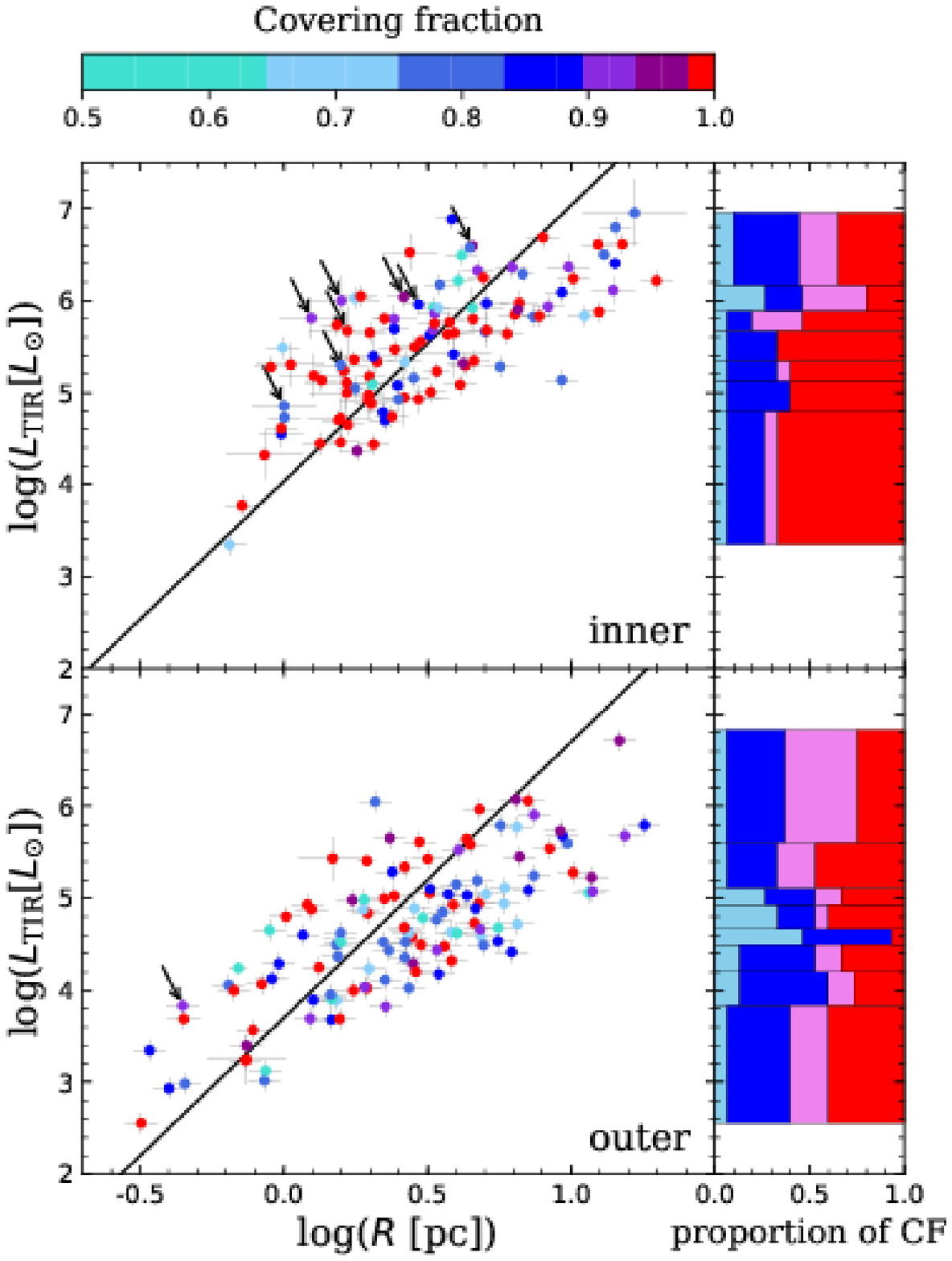}
      \end{center}
    \end{minipage}
    \hspace{5mm}
    \begin{minipage}{0.45\textwidth}
      \begin{center}
        \includegraphics[width=1.05\linewidth,bb=25 40 420 555,clip]{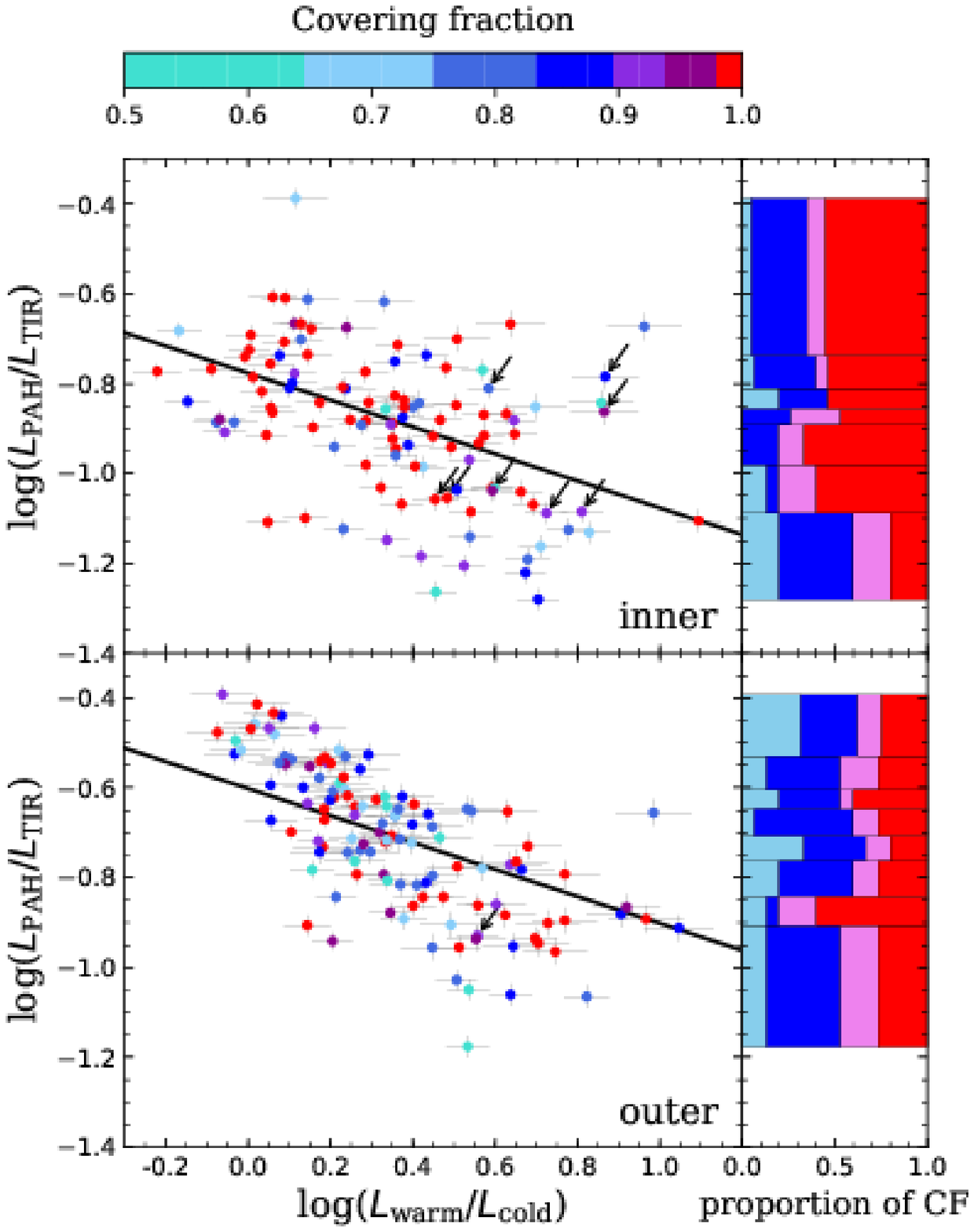}
      \end{center}
    \end{minipage}
  \end{tabular}
  \caption{
    Same relations as (left) figure~\ref{fig:Stromgren_inout} and (right) figure~\ref{fig:result_ratio}, but color-coded with the CF, and shown separately for the IR bubbles in inner ($\leq 7$~kpc) and outer ($> 7$~kpc) Galactic regions in the upper and lower panels, respectively.
    In each panel, the black line corresponds to the best-fit power-law relation to the data points of the IR bubbles with the CF of unity, where the power-law indices are fixed at 3 for the left panels while they are allowed to vary for the right panels.
    The right-hand side of each panel shows the proportion of the CF for $L_{\rm{TIR}}$ and $L_{\rm{PAH}}/L_{\rm{TIR}}$ averaged every 15 objects.
    The red, violet, blue and cyan areas correspond to CF=$1.0$, $0.9\leq$CF$<1.0$, $0.75\leq$CF$<0.9$ and $0.5\leq$CF$<0.75$, respectively.
    Only in this figure, we plot the data points of the bubbles originating from bipolar H\emissiontype{II} regions, which are denoted by arrows.
  }
  \label{fig:CF_depend}
\end{figure*}

In figure~\ref{fig:CF_depend}, we plotted the $L_{\rm{TIR}}$-$R$ and $L_{\rm{PAH}}/L_{\rm{TIR}}$-$L_{\rm{warm}}/L_{\rm{cold}}$ relations color-coded according to the CF.
In each panel, the black line corresponds to the best-fit relation for the closed (i.e., CF$=1$) bubble only.
The figure does not clearly show any systematic difference between the broken and closed bubbles, except for the $L_{\rm{TIR}}$-$R$ relation in outer Galactic regions.
The $L_{\rm{TIR}}$-$R$ relation of the broken bubbles is significantly deviated from that of the closed bubbles toward larger $R$ in outer Galactic regions.
This can be explained by considering that the IR bubbles of lower gas densities are more deviated from the $L_{\rm{TIR}}$-$R$ relation for the other IR bubbles and expected to break more easily as mentioned above. 
On the other hand, there is no systematic difference in the $L_{\rm{TIR}}$-$R$ relation between the closed and broken bubbles in inner Galactic regions, which implies that the morphology of the IR bubbles is likely to be less affected by the ambient interstellar environments in inner Galactic regions.

In order to clearly visualize trends, if any, on the right-hand side of each panel in figure~\ref{fig:CF_depend}, the proportions of the CF are shown for $L_{\rm{TIR}}$ and $L_{\rm{PAH}}/L_{\rm{TIR}}$ averaged every 15 objects.
From this figure, we can confirm the trends pointed out by \citet{Hattori2016} for the bubbles in inner Galactic regions, i.e., the IR bubbles with higher $L_{\rm{TIR}}$ and lower $L_{\rm{PAH}}/L_{\rm{TIR}}$ tend to have lower CFs.
On the other hand, we find that there is no such a trend in outer Galactic regions, and thus the trends are not universal for the bubbles along the whole Galactic plane.

Recently, \citet{Whitworth2018} developed a model for the formation of bipolar H\emissiontype{II} regions and suggested that bipolar bubbles can be formed by CCC.
As far as our bipolar bubbles (they were removed from our sample) are concerned, however, they are not necessarily attributed to CCC; the massive star associated with the bipolar H\emissiontype{II} region composed of ``EN13'' and ``EN14'' is formed by merging of multiple filaments (Mallick et al. 2013), while the massive stars associated with other two are formed by gravitational collapse of the massive clumps (``S109'', ``S110'' and ``S111''; Dalgleish et al. 2018, ``CN107'' and ``CN109''; Dewangan et al. 2016).
In figure~\ref{fig:CF_depend}, we also show their locations, from which we find that the bipolar bubbles have relatively small radii for $L_{\rm{TIR}}$ and high $L_{\rm{warm}}/L_{\rm{cold}}$ but not low $L_{\rm{PAH}}/L_{\rm{TIR}}$.
Hence those bubbles, even if included, are unlikely to contribute to the CCC trends suggested by \citet{Hattori2016}.

As another possible formation mechanism of broken morphology, we consider the champagne flow model (Tenorio-Tagle 1979).
Since a massive star forms near the edge of a cloud in this model, we expect that $L_{\rm{warm}}/L_{\rm{cold}}$ of the broken bubbles would differ systematically from that of the closed bubbles, when most of the broken bubbles are of champagne flow origin.
We performed a two sample Kolmogorov-Smirnov (K-S) test with respect to the distribution of $L_{\rm{warm}}/L_{\rm{cold}}$ between the broken (CF$< 0.9$) and closed (CF$=1$) bubbles.
The result of the K-S test shows that the distributions are not significantly different ($p>0.20$), thus not calling for the necessity of the champagne flow model to explain the broken bubbles.
To verify the possibility that the morphology and the IR properties of the bubbles may be related with the CCC mechanism, we will investigate the spatial distributions of the bubbles in the PAH and dust emissions and compare them with the CO position-velocity maps, the results of which will be reported in a separate paper.

\section{Summary}
Using AKARI and Herschel data, we obtained $R$, $L_{\rm{TIR}}$ and CF of 247 IR bubbles with $r >$\timeform{1'} along the whole Galactic plane (0$^{\circ}\leq$~l~$<$360$^{\circ}$, $|$b$|\leq$~5$^{\circ}$).
As a result, we find that there are systematic differences in the properties of the IR bubbles between the inner and outer Galactic regions; $L_{\rm{TIR}}$ and CF are systematically lower while $L_{\rm{PAH}}/L_{\rm{TIR}}$ is higher in outer Galactic regions.
Investigating the dependence of those properties on the Galactocentric distance, we suggest that the results are explained by the changes in the properties of the massive stars and the interstellar environments associated with the Galactic IR bubbles from inner to outer Galactic regions; (1) the central massive stars are likely to be of later spectral types, (2) the IR bubbles may be more easily expanded and broken due to lower ambient interstellar pressure and (3) the higher ratios of C-rich AGBs to O-rich AGBs may cause the shells of the IR bubbles to be richer in PAH in outer Galactic regions.
Finally, the IR bubbles in outer Galactic regions do not show evidence for the possibility that large broken IR bubbles may be caused by CCC, although the IR bubbles in inner Galactic regions show results consistent with those in \citet{Hattori2016} which are indicative of CCC for large broken IR bubbles.

\begin{ack}
  We thank the referee for carefully reading our manuscript and giving us helpful comments.
  This research is based on observations with AKARI, a JAXA project with the participation of ESA.
  Herschel is an ESA space observatory with science instruments provided by European-led Principal Investigator consortia and with important participation from NASA.
  We thank all the members of the AKARI and Herschel projects, particularly the all-sky survey and Hi-GAL data reduction teams.
\end{ack}

\appendix
\section*{Data set of the IR bubbles discussed in this study.}
\begin{longtable}{p{0.5cm} r @{\hspace{-0.2cm}} r @{\hspace{-0.2cm}} r @{\hspace{0.7cm}} c c c @{\hspace{-0.3cm}} r @{\hspace{0.7cm}} p{1.2cm}}
  \caption{Summary of the central positions, radii, distances and covering fractions of the IR bubbles.}
  \label{longtable:l_b}
  \hline
  Name & $l$ [\timeform{D}] & $b$ [\timeform{D}] & $r$ [\timeform{'}] & $D$ [kpc] & Distance ref.$\dagger$ & $D_{\rm{Gal.}}$ [kpc] & $R$ [pc] & CF \\
  \endfirsthead
  \hline
  Name & $l$ [\timeform{D}] & $b$ [\timeform{D}] & $r$ [\timeform{'}] & $D$ [kpc] & Distance ref.$\dagger$ & $D_{\rm{Gal.}}$ [kpc] & $R$ [pc] & CF \\
  \hline
  \endhead
  \hline
  \endfoot
  \hline
  \multicolumn{9}{p{150mm}}{\small\slshape \strut $\dagger$References: Beaumont+10, \citet{Beaumont2010}; Churchwell+06, \citet{Churchwell2006}; Churchwell+07, \citet{Churchwell2007}; Deharveng+09, \citet{Deharveng2009}; Deharveng+10, \citet{Deharveng2010}; Dewangan+15, \citet{Dewangan2015}; Gennaro+12, \citet{Gennaro2012}; Hou\&Gao+14, \citet{Hou&Gao2014}; Hou\&Han+14, \citet{Hou2014}; Pavel+12, \citet{Pavel2012}; Rahman+10, \citet{Rahman2010}; Rodr{\'{\i}}guez+12, \citet{Rodriguez2012}; Watson+09, \citet{Watson2009}; Watson+10, \citet{Watson2010}; Zhang+13, \citet{Zhang2013}; Orion, Perseus, Sagittarius, the distance is estimated by assuming the IR bubble is located on each Galactic spiral arm. \newline $\S$ The nearer kinematic distance is adopted.\strut}\\
  \endlastfoot
  \hline
N2	&	$10.749$	&	$-0.468$	&	$6.81$	&	8.4$\pm$3.5	&	Churchwell+06	&	1.6$\pm$3.5	&	$16.63$	&	$0.75$	\\
N4	&	$11.892$	&	$0.750$	&	$2.01$	&	3.4$\pm$0.4	&	Watson+10	&	5.2$\pm$0.4	&	$1.99$	&	$1.00$	\\
N5	&	$12.478$	&	$-1.140$	&	$3.63$	&	3.7$\pm$1.1	&	Beaumont+10	&	5.0$\pm$1.1	&	$3.90$	&	$0.83$	\\
N6	&	$12.522$	&	$-0.625$	&	$6.18$	&	4.1$\pm$0.5$\S$	&	Deharveng+10	&	4.6$\pm$0.5	&	$7.40$	&	$0.75$	\\
N10	&	$13.188$	&	$0.040$	&	$1.49$	&	4.6$\pm$0.6	&	Deharveng+10	&	4.2$\pm$0.6	&	$1.99$	&	$1.00$	\\
N11	&	$13.225$	&	$0.085$	&	$1.11$	&	3.8$\pm$0.5	&	Watson+10	&	4.9$\pm$0.5	&	$1.23$	&	$1.00$	\\
N12	&	$13.724$	&	$-0.018$	&	$4.98$	&	4.4$\pm$0.6$\S$	&	Deharveng+10	&	4.4$\pm$0.6	&	$6.37$	&	$1.00$	\\
N14	&	$14.004$	&	$-0.142$	&	$1.00$	&	3.1$\pm$0.4	&	Watson+10	&	5.5$\pm$0.4	&	$0.90$	&	$1.00$	\\
\end{longtable}

\begin{landscape}{
    {\renewcommand{\arraystretch}{1.4}%
  \fontsize{8pt}{8pt}\selectfont
  \setlength{\textwidth}{26cm}
  \setlength{\tabcolsep}{2pt}
  \begin{longtable}[htb]{ccccccccccccc}
    \caption{Summary of the aperture-corrected flux densities of the IR bubbles obtained by the aperture photometry.}
    \label{longtable:flux}
    \hline
    Name & Flux 9 $\mu$m [Jy] & Flux 18 $\mu$m [Jy] & Flux 65 $\mu$m [Jy] & Flux 90 $\mu$m [Jy] & Flux 140 $\mu$m [Jy] & Flux 160 $\mu$m [Jy] & Flux 70 $\mu$m [Jy] & Flux 160 $\mu$m [Jy] & Flux 250 $\mu$m [Jy] & Flux 350 $\mu$m [Jy] & Flux 500 $\mu$m [Jy] & Notes$\dagger$\\
     & (AKARI IRC) & (AKARI IRC) & (AKARI FIS) & (AKARI FIS) & (AKARI FIS) & (AKARI FIS) & (Herschel PACS) & (Herschel PACS) & (Herschel SPIRE) & (Herschel SPIRE) & (Herschel SPIRE) &\\
    \endfirsthead
    \hline
    Name & Flux 9 $\mu$m [Jy] & Flux 18 $\mu$m [Jy] & Flux 65 $\mu$m [Jy] & Flux 90 $\mu$m [Jy] & Flux 140 $\mu$m [Jy] & Flux 160 $\mu$m [Jy] & Flux 70 $\mu$m [Jy] & Flux 160 $\mu$m [Jy] & Flux 250 $\mu$m [Jy] & Flux 350 $\mu$m [Jy] & Flux 500 $\mu$m [Jy] & Notes$\dagger$\\
     & (AKARI IRC) & (AKARI IRC) & (AKARI FIS) & (AKARI FIS) & (AKARI FIS) & (AKARI FIS) & (Herschel PACS) & (Herschel PACS) & (Herschel SPIRE) & (Herschel SPIRE) & (Herschel SPIRE) & \\
    \hline
    \endhead
    \hline
    \endfoot
    \hline
    \multicolumn{13}{l}{We considered random and systematic errors as flux uncertainties (see text for detail).}\\
    \multicolumn{13}{l}{$\dagger$sc: Targets with source contamination in their background regions, where areas larger than 10\% suffer source contamination.}\\
    \endlastfoot
    \hline
N2	&	(8.6$\pm$0.9)$\times10^2$	&	(1.2$\pm$0.1)$\times10^3$	&	(3.0$\pm$0.4)$\times10^4$	&	(4.0$\pm$0.6)$\times10^4$	&	(5.3$\pm$0.8)$\times10^4$	&	(7.0$\pm$1.0)$\times10^4$	&	(3.5$\pm$0.4)$\times10^4$	&	(7.1$\pm$0.7)$\times10^4$	&	(3.9$\pm$0.4)$\times10^4$	&	(1.5$\pm$0.2)$\times10^4$	&	(5.7$\pm$0.6)$\times10^3$	&	$-$	\\
N4	&	(1.5$\pm$0.1)$\times10^2$	&	(1.6$\pm$0.2)$\times10^2$	&	(3.2$\pm$0.5)$\times10^3$	&	(3.6$\pm$0.5)$\times10^3$	&	(3.8$\pm$0.6)$\times10^3$	&	(5.0$\pm$0.8)$\times10^3$	&	(3.3$\pm$0.3)$\times10^3$	&	(4.7$\pm$0.5)$\times10^3$	&	(2.3$\pm$0.2)$\times10^3$	&	(8.6$\pm$0.9)$\times10^2$	&	(3.1$\pm$0.3)$\times10^2$	&	$-$	\\
N5	&	(1.7$\pm$0.2)$\times10^2$	&	(2.3$\pm$0.2)$\times10^2$	&	(3.8$\pm$0.6)$\times10^3$	&	(5.5$\pm$0.8)$\times10^3$	&	(9.0$\pm$1.0)$\times10^3$	&	(9.0$\pm$1.0)$\times10^3$	&	$-$	&	$-$	&	$-$	&	$-$	&	$-$	&	$-$	\\
N6	&	(4.6$\pm$0.5)$\times10^2$	&	(4.6$\pm$0.5)$\times10^2$	&	(8.0$\pm$1.0)$\times10^3$	&	(1.1$\pm$0.2)$\times10^4$	&	(1.9$\pm$0.3)$\times10^4$	&	(2.1$\pm$0.3)$\times10^4$	&	(8.6$\pm$0.9)$\times10^3$	&	(2.0$\pm$0.2)$\times10^4$	&	(1.2$\pm$0.1)$\times10^4$	&	(4.8$\pm$0.5)$\times10^3$	&	(1.9$\pm$0.2)$\times10^3$	&	$-$	\\
N10	&	(1.2$\pm$0.1)$\times10^2$	&	(3.4$\pm$0.3)$\times10^2$	&	(6.5$\pm$1.0)$\times10^3$	&	(5.2$\pm$0.8)$\times10^3$	&	(7.0$\pm$1.0)$\times10^3$	&	(8.0$\pm$1.0)$\times10^3$	&	(5.0$\pm$0.5)$\times10^3$	&	(5.9$\pm$0.6)$\times10^3$	&	(3.5$\pm$0.3)$\times10^3$	&	(1.5$\pm$0.2)$\times10^3$	&	(6.1$\pm$0.6)$\times10^2$	&	$-$	\\
N11	&	(4.5$\pm$0.5)$\times10^1$	&	(5.3$\pm$0.5)$\times10^1$	&	(1.9$\pm$0.3)$\times10^3$	&	(2.0$\pm$0.3)$\times10^3$	&	(2.6$\pm$0.4)$\times10^3$	&	(3.1$\pm$0.5)$\times10^3$	&	(1.2$\pm$0.1)$\times10^3$	&	(1.3$\pm$0.1)$\times10^3$	&	(6.0$\pm$0.6)$\times10^2$	&	(2.6$\pm$0.3)$\times10^2$	&	(9.0$\pm$1.0)$\times10^1$	&	$-$	\\
\end{longtable}}
}\end{landscape}

\begin{longtable}{ccccc}
  \caption{Summary of the luminosities of the IR bubbles obtained by the SED fitting.}
  \label{longtable:luminosity}
  \hline
  Name & log($L_{\rm{TIR}}/$L$_{\odot}$) & log($L_{\rm{PAH}}/L_{\rm{TIR}}$) & log($L_{\rm{warm}}/L_{\rm{TIR}}$) & log($L_{\rm{cold}}/L_{\rm{TIR}}$)\\
  \endfirsthead
  \hline
  Name & log($L_{\rm{TIR}}/$L$_{\odot}$) & log($L_{\rm{PAH}}/L_{\rm{TIR}}$) & log($L_{\rm{warm}}/L_{\rm{TIR}}$) & log($L_{\rm{cold}}/L_{\rm{TIR}}$)\\
  \hline
  \endhead
  \hline
  \endfoot
  \hline
  \endlastfoot
  \hline
N2	&	6.95$\pm$0.36	&	$-0.94\pm$0.02	&	$-0.26\pm$0.04	&	$-0.47\pm$0.03	\\
N4	&	5.17$\pm$0.11	&	$-0.71\pm$0.02	&	$-0.25\pm$0.04	&	$-0.61\pm$0.03	\\
N5	&	5.41$\pm$0.26	&	$-0.81\pm$0.03	&	$-0.27\pm$0.05	&	$-0.51\pm$0.05	\\
N6	&	5.82$\pm$0.11	&	$-0.70\pm$0.02	&	$-0.34\pm$0.04	&	$-0.47\pm$0.03	\\
N10	&	5.65$\pm$0.12	&	$-1.03\pm$0.02	&	$-0.14\pm$0.04	&	$-0.73\pm$0.03	\\
N12	&	5.85$\pm$0.11	&	$-0.76\pm$0.02	&	$-0.36\pm$0.05	&	$-0.41\pm$0.03	\\
N14	&	5.28$\pm$0.12	&	$-0.87\pm$0.02	&	$-0.16\pm$0.04	&	$-0.78\pm$0.03	\\
N16	&	6.09$\pm$0.11	&	$-0.84\pm$0.02	&	$-0.45\pm$0.05	&	$-0.30\pm$0.03	\\
\end{longtable}

\clearpage
\begin{figure*}[ht]
  \begin{center}
    \subfigure{
      \mbox{\raisebox{0mm}{\includegraphics[width=180mm, bb=30 40 600 400, clip]{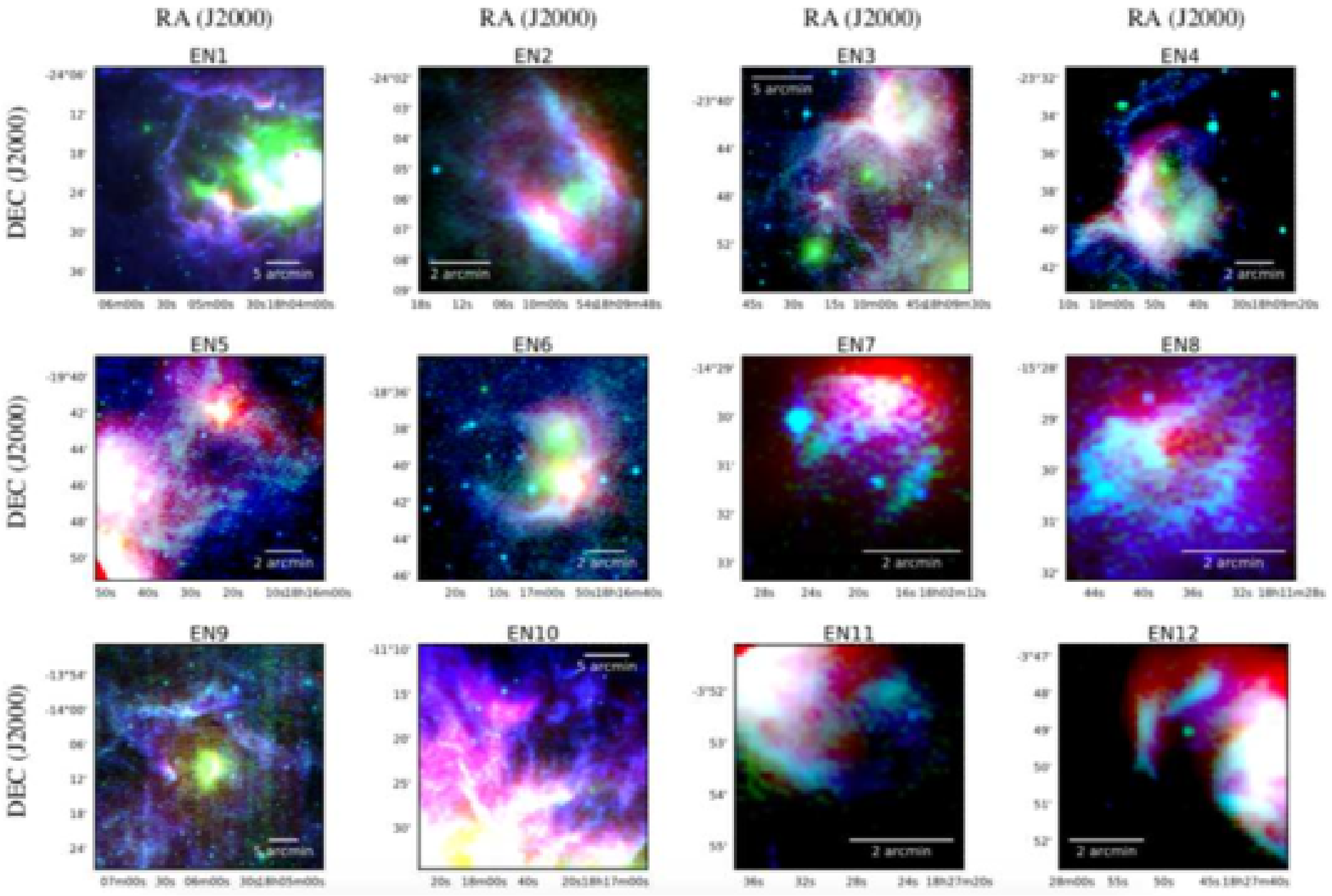}}}
    }
    \subfigure{
      \mbox{\raisebox{0mm}{\includegraphics[width=180mm, bb=30 40 610 390, clip]{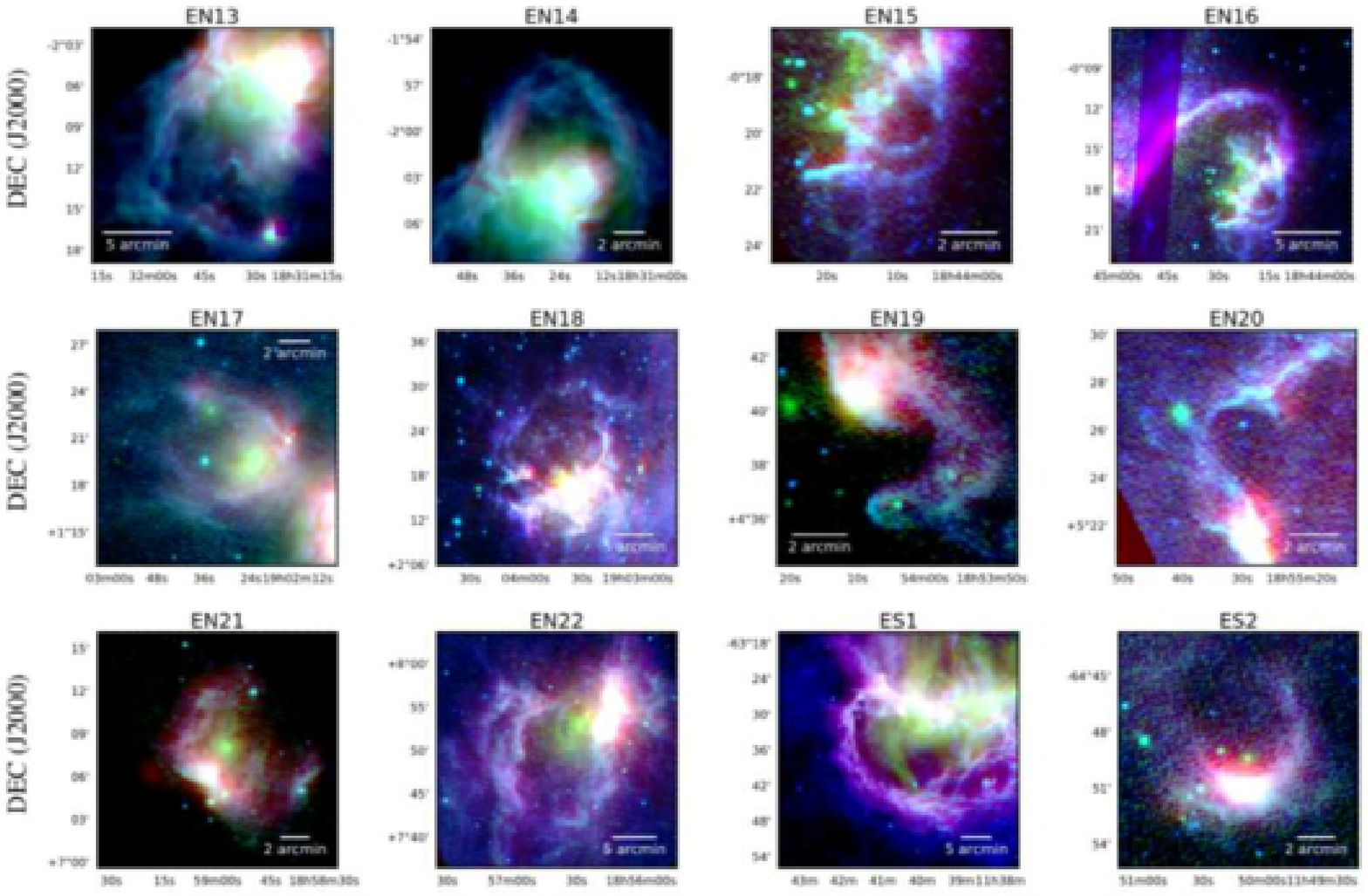}}}
    }
  \end{center}
  \caption{Images of the IR bubbles newly found in this study. The blue, green and red correspond to the AKARI 9 $\mu$m, 18 $\mu$m and 90 $\mu$m band images, respectively.}
\end{figure*}
\addtocounter{figure}{-1}
\begin{figure*}[ht]
  \begin{center}
    \subfigure{
      \mbox{\raisebox{0mm}{\includegraphics[width=180mm, bb=30 40 600 400, clip]{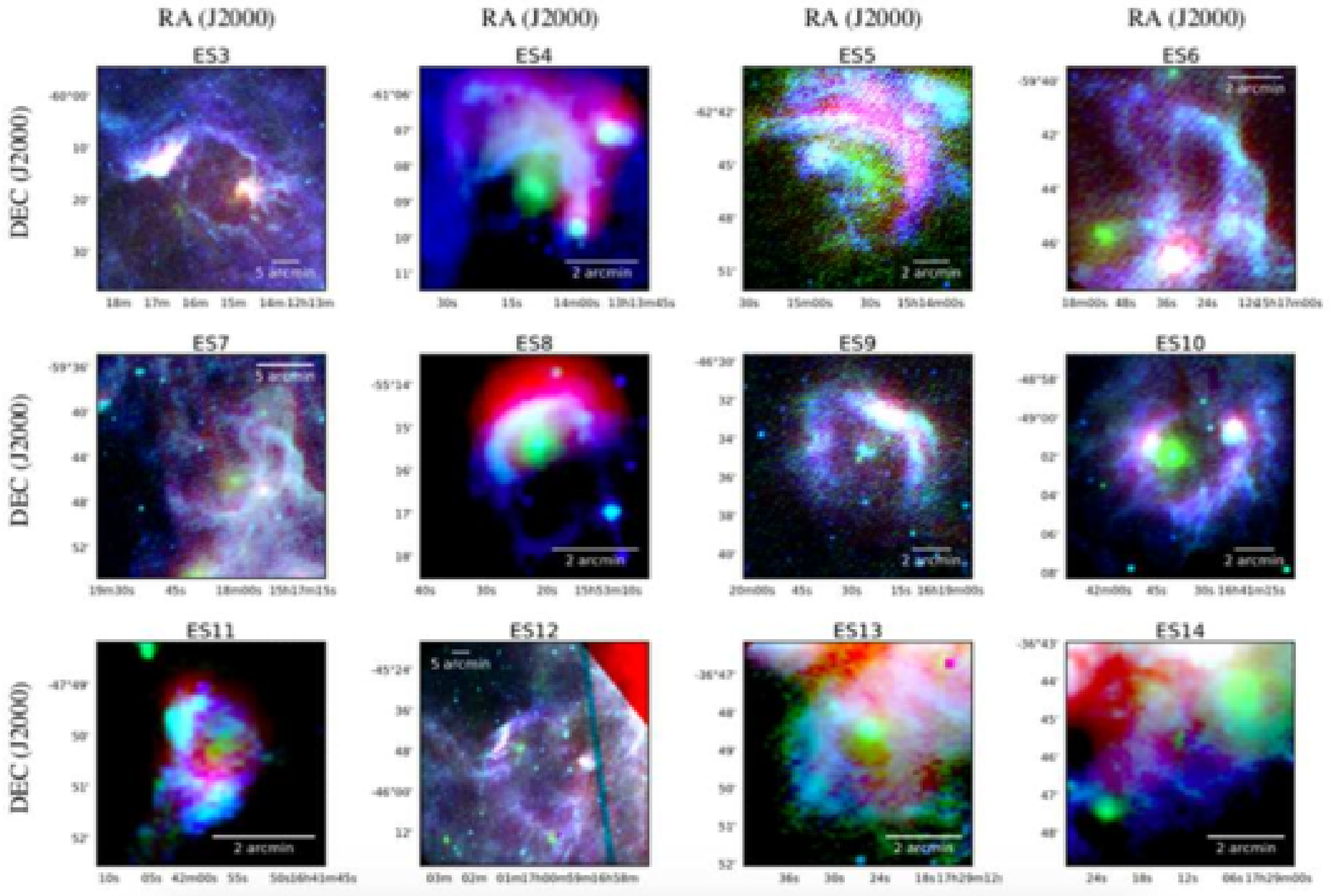}}}
    }
    \subfigure{
      \mbox{\raisebox{0mm}{\includegraphics[width=180mm, bb=30 40 600 390, clip]{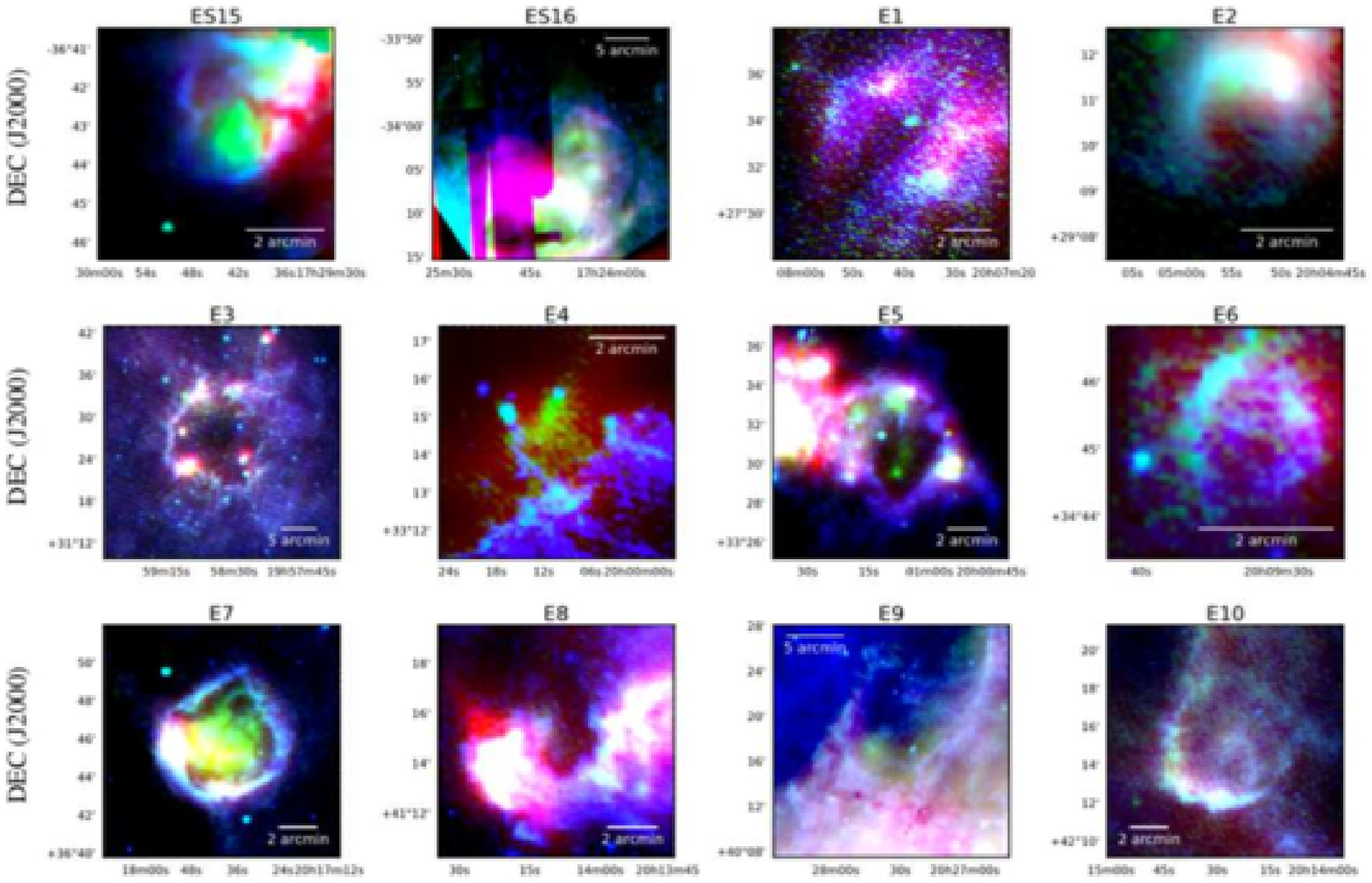}}}
    }
  \end{center}
  \caption{Continued.}
\end{figure*}
\addtocounter{figure}{-1}
\begin{figure*}[ht]
  \begin{center}
    \subfigure{
      \mbox{\raisebox{0mm}{\includegraphics[width=180mm, bb=30 40 600 400, clip]{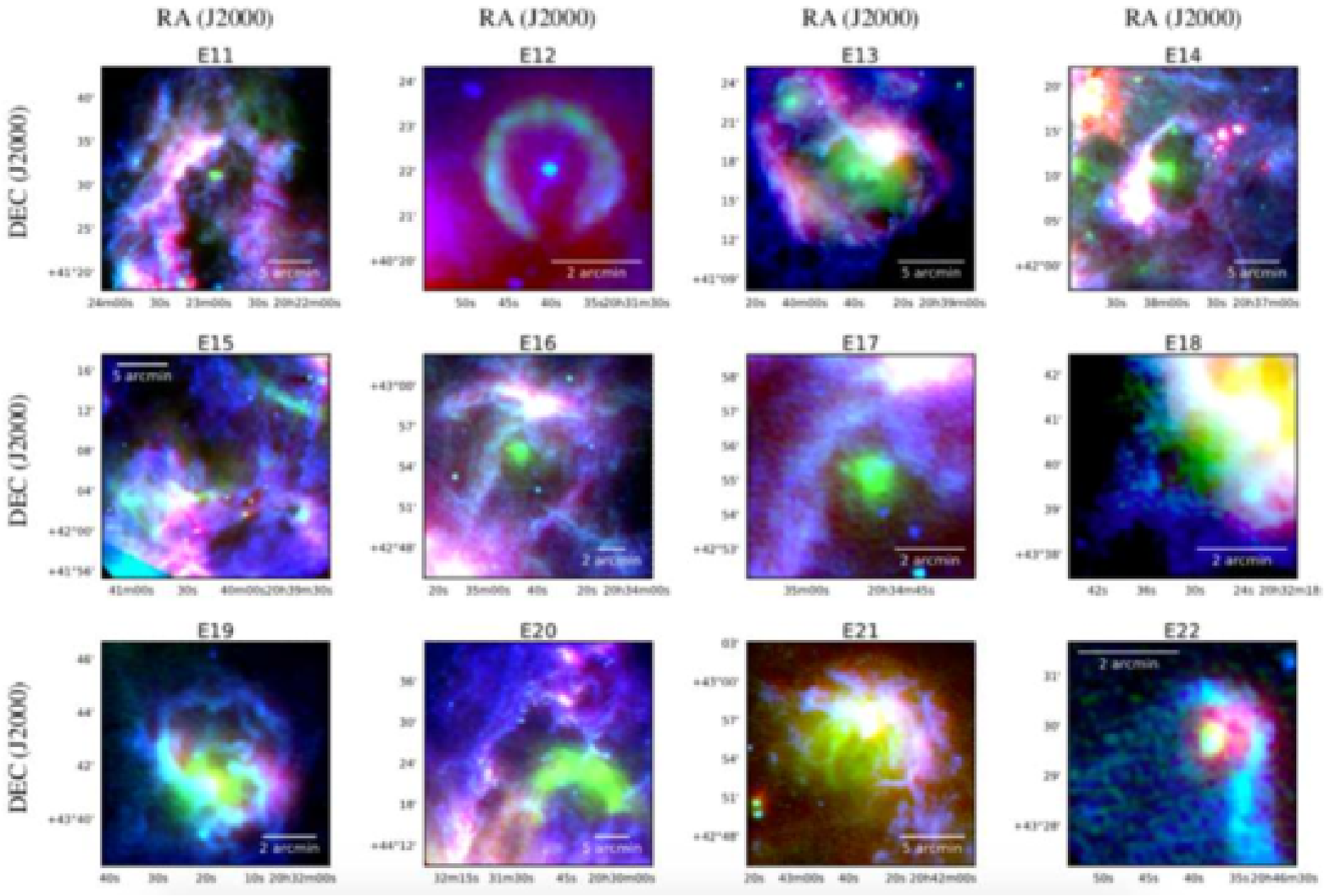}}}
    }
    \subfigure{
      \mbox{\raisebox{0mm}{\includegraphics[width=180mm, bb=30 40 600 390, clip]{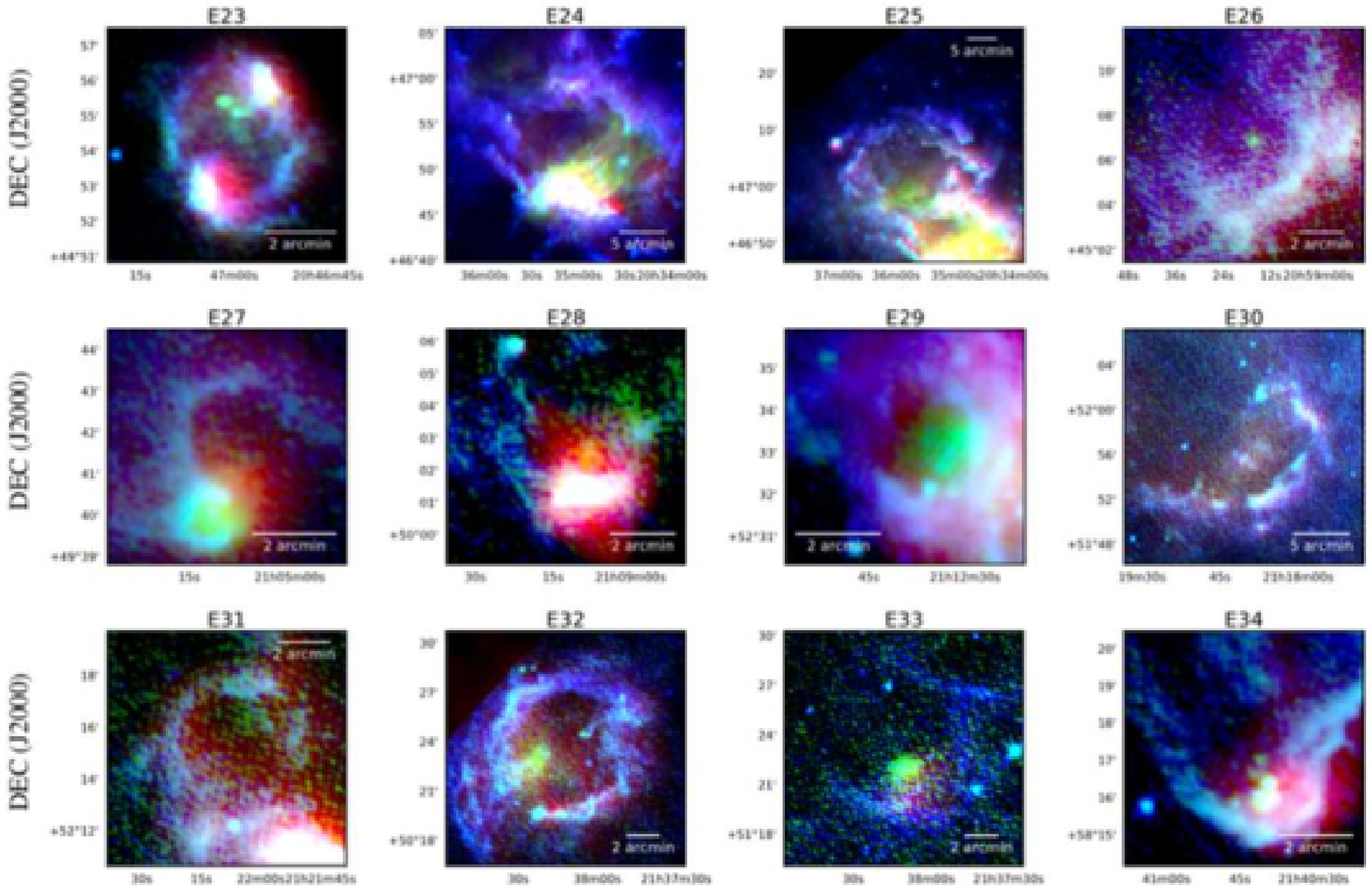}}}
    }
  \end{center}
  \caption{Continued.}
\end{figure*}
\addtocounter{figure}{-1}
\begin{figure*}[ht]
  \begin{center}
    \subfigure{
      \mbox{\raisebox{0mm}{\includegraphics[width=180mm, bb=30 40 600 400, clip]{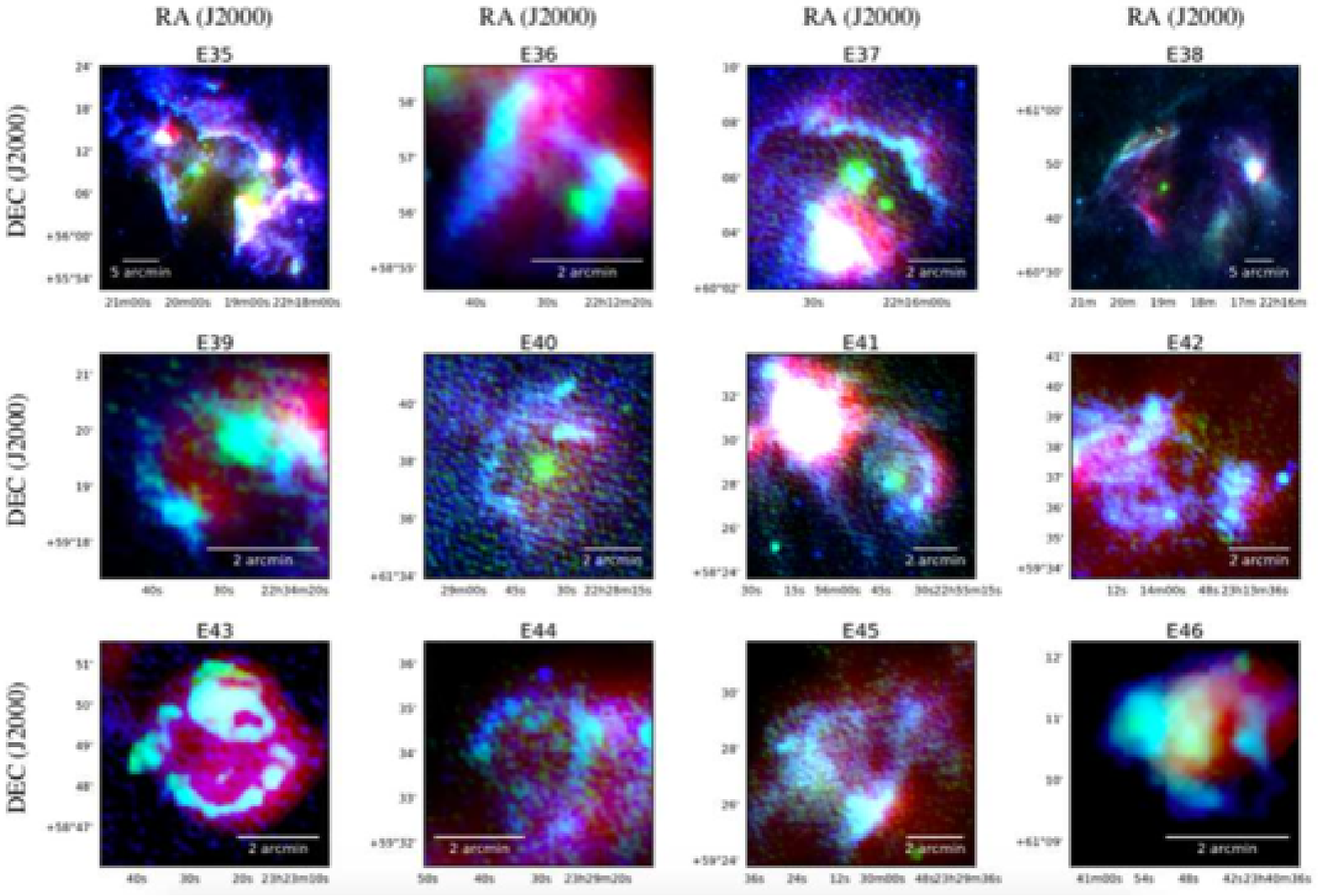}}}
    }
    \subfigure{
      \mbox{\raisebox{0mm}{\includegraphics[width=180mm, bb=30 40 600 390, clip]{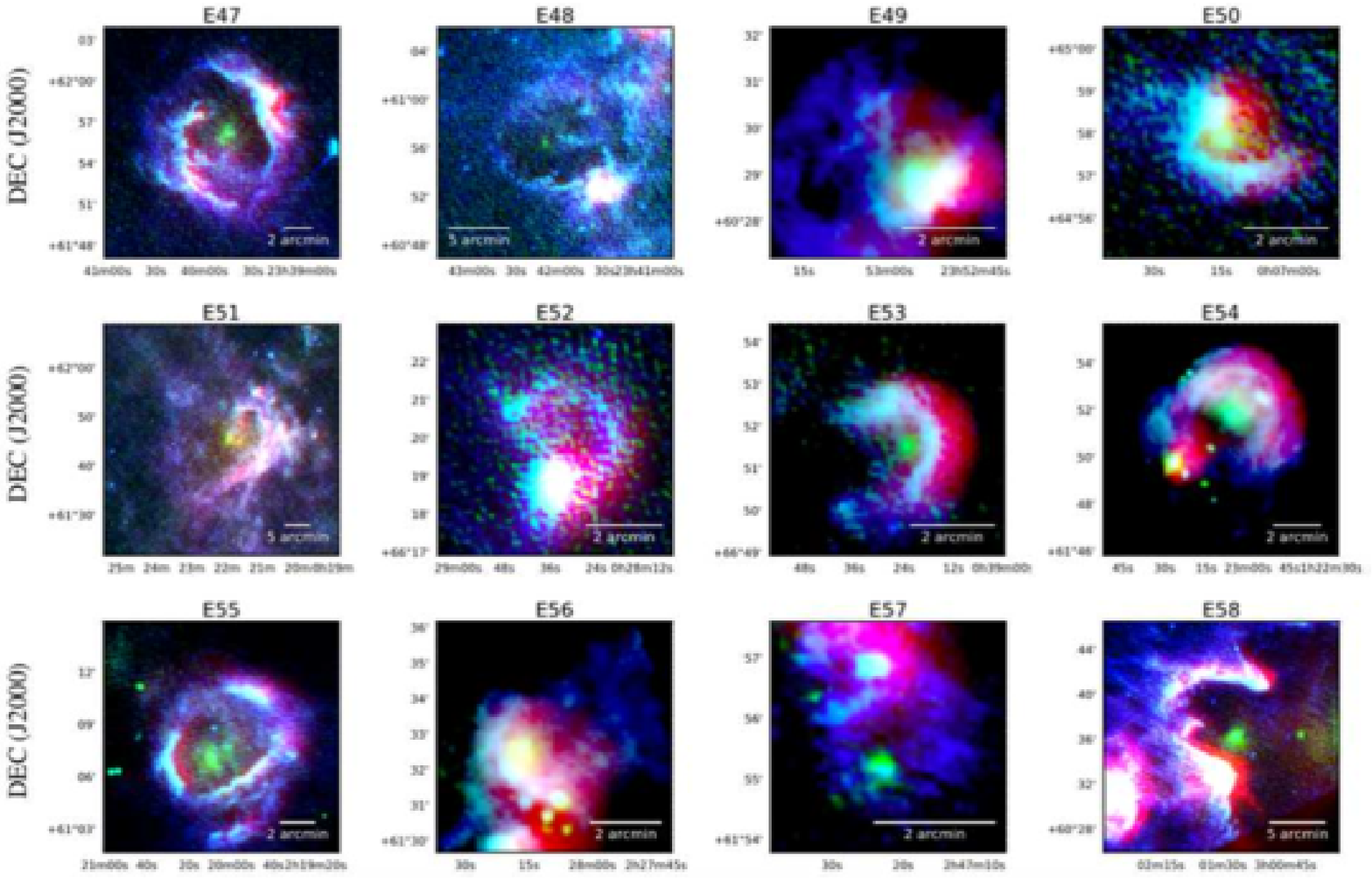}}}
    }
  \end{center}
  \caption{Continued.}
\end{figure*}
\addtocounter{figure}{-1}
\begin{figure*}[ht]
  \begin{center}
    \subfigure{
      \mbox{\raisebox{0mm}{\includegraphics[width=180mm, bb=30 40 600 400, clip]{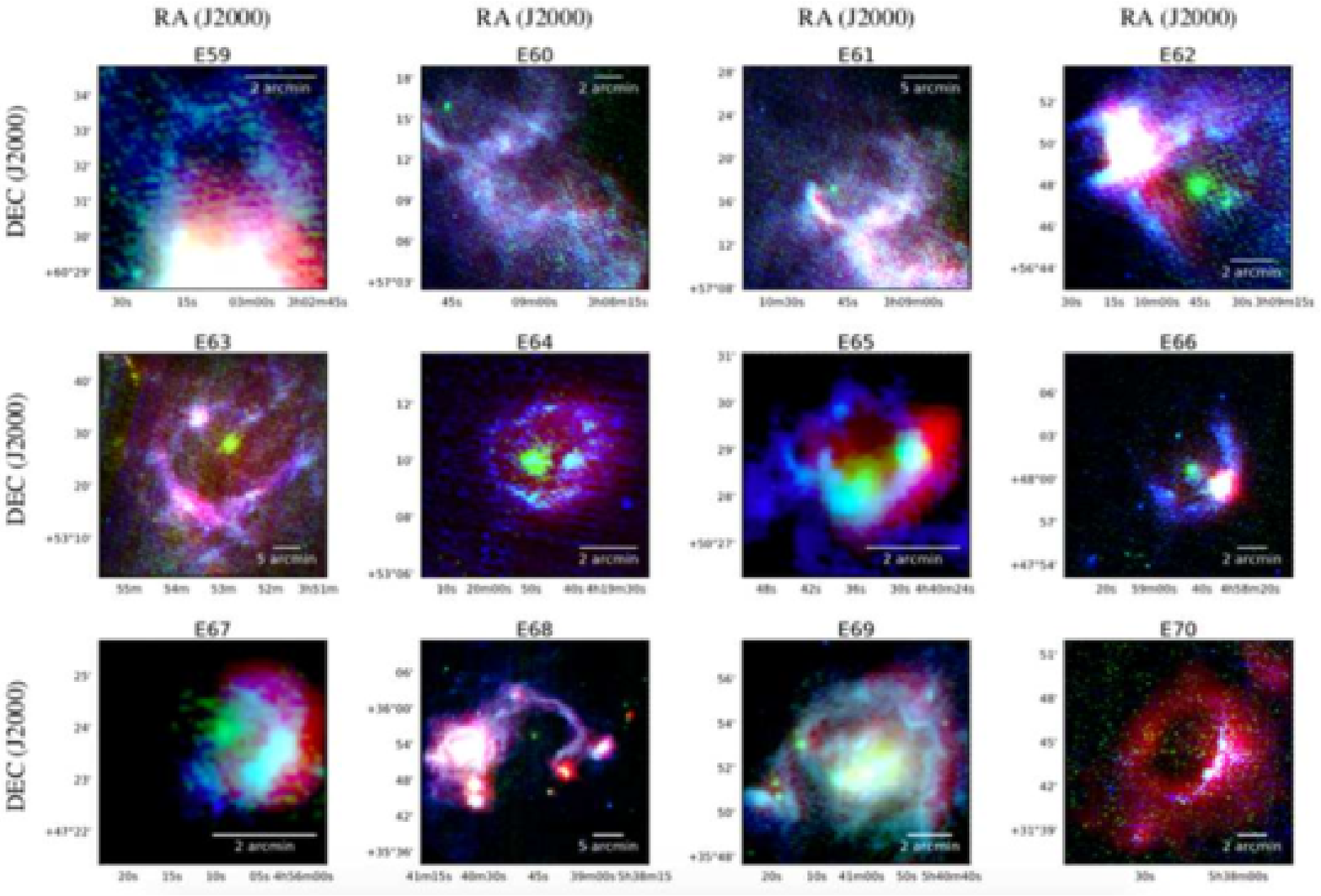}}}
    }
    \subfigure{
      \mbox{\raisebox{0mm}{\includegraphics[width=180mm, bb=30 40 600 390, clip]{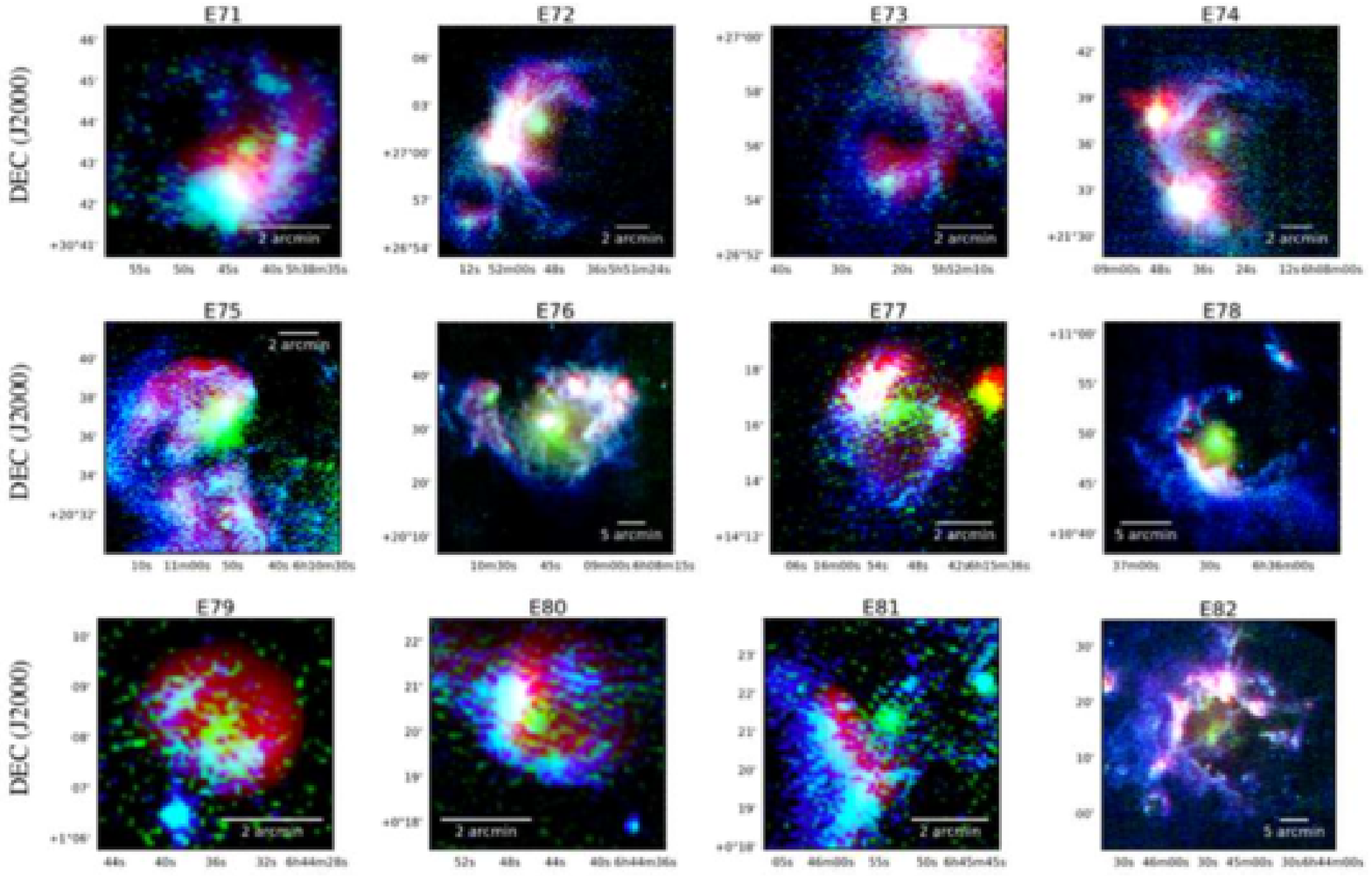}}}
    }
  \end{center}
  \caption{Continued.}
\end{figure*}
\addtocounter{figure}{-1}
\begin{figure*}[ht]
  \begin{center}
    \subfigure{
      \mbox{\raisebox{0mm}{\includegraphics[width=180mm, bb=30 40 600 400, clip]{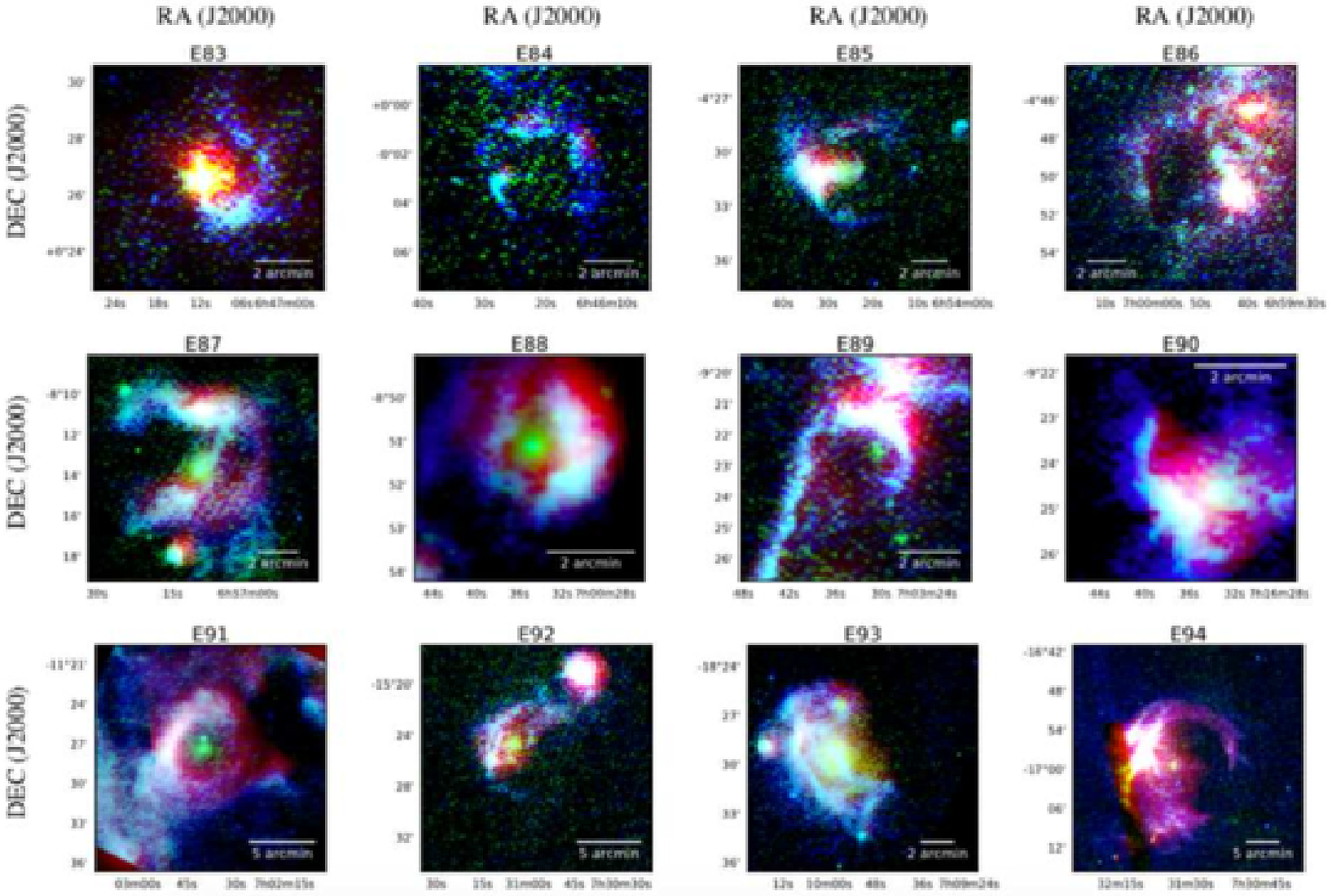}}}
    }
    \subfigure{
      \mbox{\raisebox{0mm}{\includegraphics[width=180mm, bb=30 40 600 390, clip]{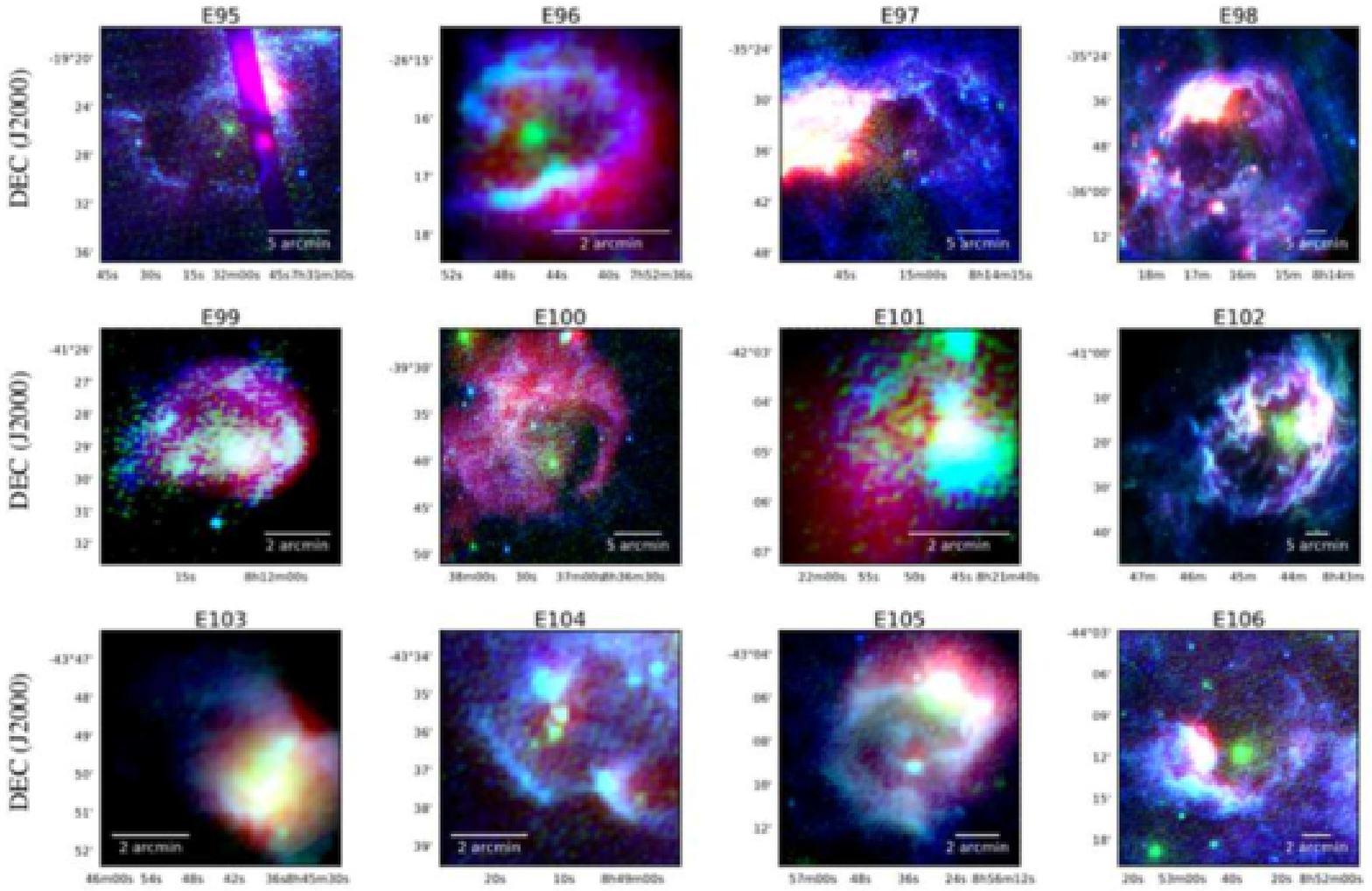}}}
    }
  \end{center}
  \caption{Continued.}
\end{figure*}
\addtocounter{figure}{-1}
\begin{figure*}[ht]
  \begin{center}
    \subfigure{
      \mbox{\raisebox{0mm}{\includegraphics[width=180mm, bb=30 40 600 400, clip]{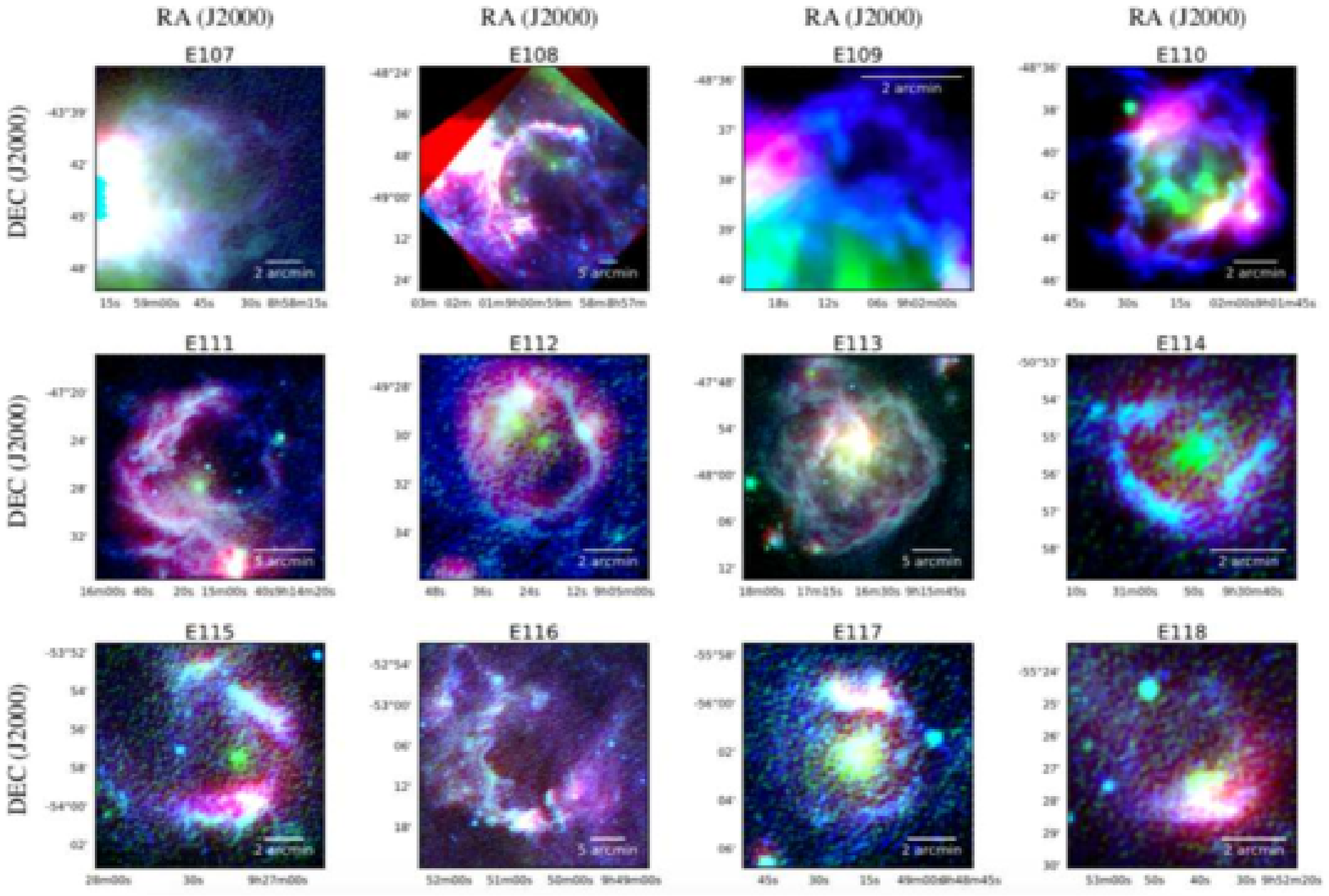}}}
    }
    \subfigure{
      \mbox{\raisebox{0mm}{\includegraphics[width=180mm, bb=30 40 600 390, clip]{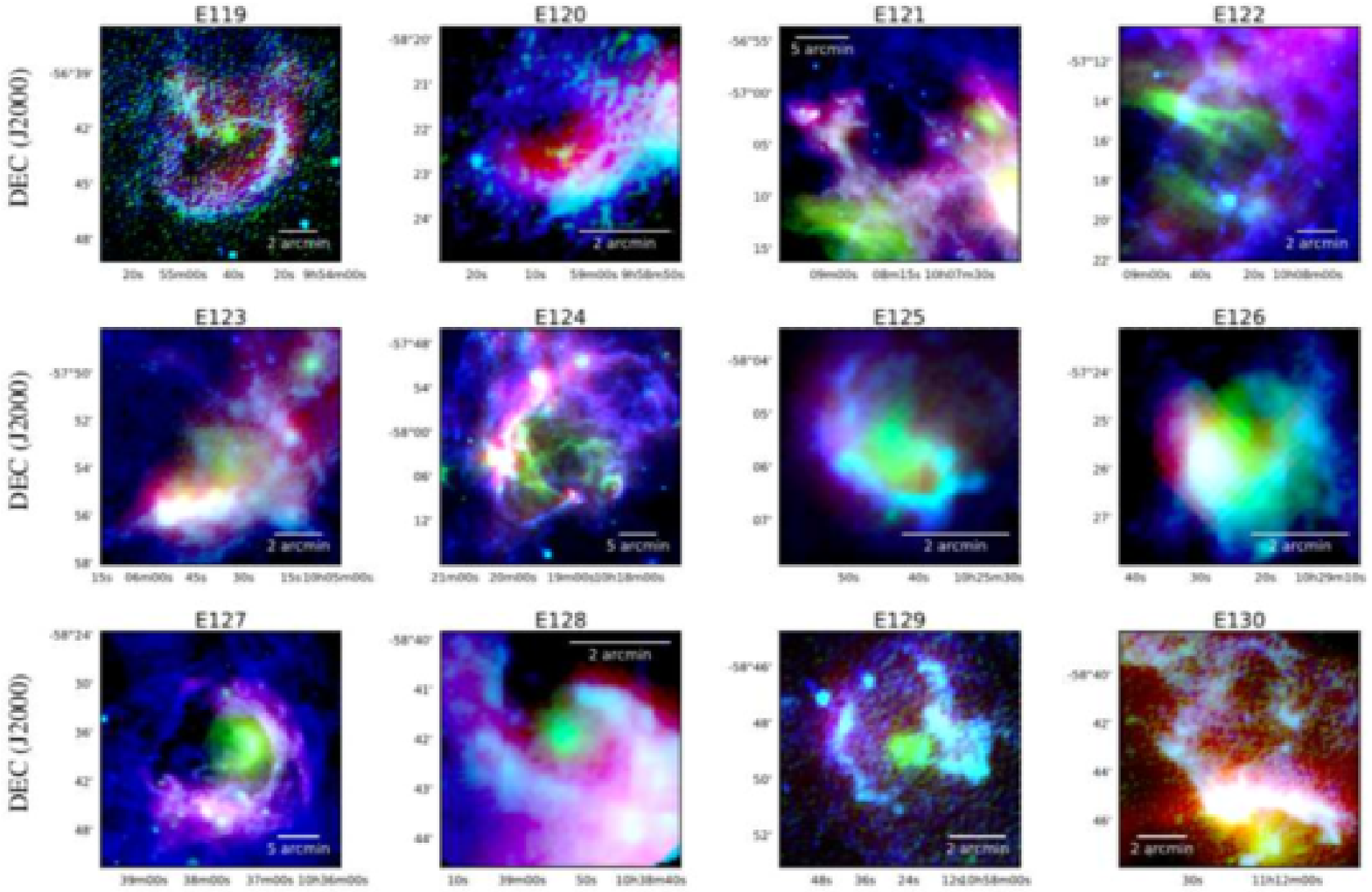}}}
    }
  \end{center}
  \caption{Continued.}
\end{figure*}
\addtocounter{figure}{-1}
\begin{figure*}[ht]
  \begin{center}
    \subfigure{
      \mbox{\raisebox{0mm}{\includegraphics[width=180mm, bb=30 40 600 400, clip]{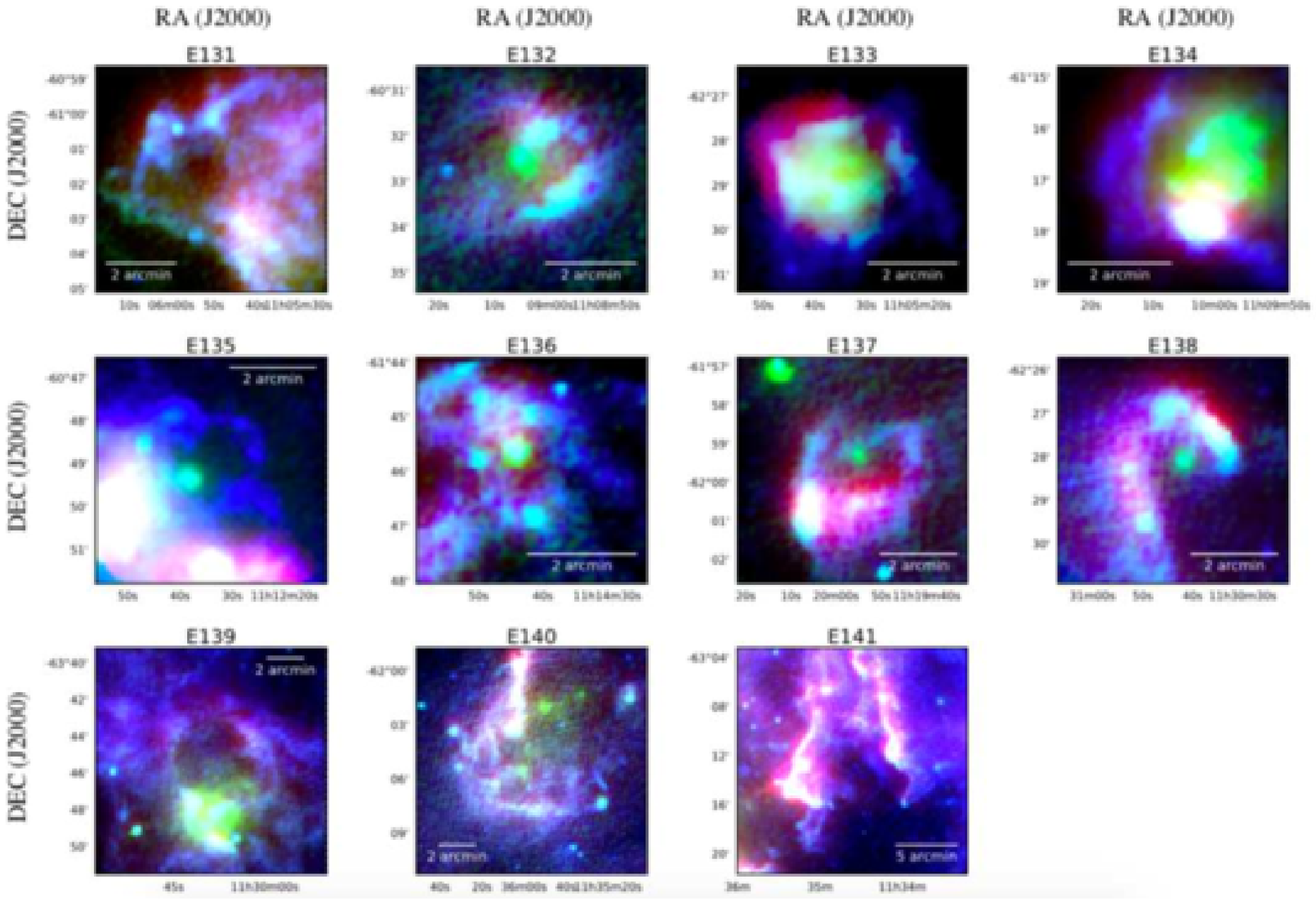}}}
    }
  \end{center}
  \caption{Continued.}
\end{figure*}

\clearpage



\begin{thebibliography}{}
\bibitem[Acero et al.(2016)]{Acero2016}
  Acero, F., et al. 2016, ApJS, 224, 8
\bibitem[Ackermann et al.(2012)]{Ackermann2012}
  Ackermann, M., et al. 2012, \apj, 750, 3
\bibitem[Arendt et al.(1998)]{Arendt1998}
  Arendt, R.G., et al. 1998, \apj, 508, 74
\bibitem[Anderson et al.(2012)]{Anderson2012}
  Anderson, L.D., et al. 2012, A\&A, 542, A10


\bibitem[Bally, Snell and Predmore (1983)]{Bally1983}
  Bally, J., Snell, R.L., \& Predmore, R. 1983, ApJ, 272, 154
\bibitem[Bassino et al.(1982)]{Bassino1982}
  Bassino, L.P., Dessaunet, V.H., Muzzio, J.C., \& Waldhausen, S. 1982, \mnras, 201, 885
\bibitem[Baug et al.(2016)]{Baug2016}
  Baug, T., Dewangan, L.K., Ojha, D.K., \& Ninan, J.P. 2016, \apj, 833, 85
\bibitem[Beaumont and Williams(2010)]{Beaumont2010}
  Beaumont, C.N., \& Williams, J.P. 2010, \apj, 709, 791
\bibitem[Benjamin et al.(2003)]{Benjamin2003}  
  Benjamin, R.A., et al. 2003, \pasp, 115, 953
\bibitem[Bronfman et al.(2000)]{Bronfman2000}
  Bronfman, L., Casassus, S., May, J., \& Nyman, L. 2000, \aap, 358, 521

\bibitem[Case and Bhattacharya(1998)]{Case1998}
  Case, G.L., \& Bhattacharya, D. 1998, \apj, 504, 761
\bibitem[Churchwell et al.(2006)]{Churchwell2006}
  Churchwell, E., et al. 2006, \apj, 649, 759
\bibitem[Churchwell et al.(2007)]{Churchwell2007}
  Churchwell, E., et al. 2007, \apj, 670, 428
\bibitem[Churchwell et al.(2009)]{Churchwell2009}
  Churchwell, E., et al. 2009, \pasp, 121, 213

\bibitem[Dale, Bonnell and Whitworth(2007)]{Dale2007}
  Dale, J.E., Bonnell, I.A., \& Whitworth, A.P. 2007, \mnras, 375, 1291
\bibitem[Dalgleish et al.(2018)]{Dalgleish2018}
  Dalgleish, H.S., Longmore, S.N., Peters, T., Henshaw, J.D., Veitch-Michaelis, J.L., \& Urquhart, J.S. 2018, \mnras, 478, 3530
\bibitem[Deharveng et al.(2009)]{Deharveng2009}  
  Deharveng, L., Zavagno, A., Schuller, F., Caplan, J., Pomar{\`e}s, M., \& Breuck, C.D. 2009, \aap, 496, 177
\bibitem[Deharveng et al.(2010)]{Deharveng2010}  
  Deharveng, L., et al. 2010, \aap, 523, A6
\bibitem[Deharveng et al.(2012)]{Deharveng2012}  
  Deharveng, L., et al. 2012, \aap, 546, A74
\bibitem[Deharveng et al.(2015)]{Deharveng2015}
  Deharveng, L., et al. 2015, \aap, 582, A1
\bibitem[Dewangan et al.(2015)]{Dewangan2015}  
  Dewangan, L.K., Ojha, D.K., Grave, J.M.C., \& Mallick, K.K. 2015, \mnras, 446, 2640
\bibitem[Dewangan et al.(2016)]{Dewangan2016}  
  Dewangan, L.K., Ojha, D.K., Zinchenko, I., Janardhan, P., Ghosh, S.K., \& Luna, A. 2016, \apj, 833, 246
\bibitem[Doi et al.(2015)]{Doi2015}
  Doi, Y., et al. 2015, \pasj, 67, 50
\bibitem[Draine(2003)]{Draine2003a}
  Draine, B.T. 2003, \araa, 41, 241
\bibitem[Draine and Li(2007)]{Draine&Li2007}
  Draine, B.T., \& Li, A. 2007, \apj, 657, 810
\bibitem[Draine(2011)]{Draine2011}
  Draine, B.T. 2011, Physics of the Interstellar and Intergalactic Medium, Princeton University Press
\bibitem[Dubner et al.(1992)]{Dubner1992}
  Dubner, G., Giacani, E., Cappa de Nicolau, C., \& Reynoso, E. 1992, A\&AS, 96, 505

\bibitem[Elmegreen(1998)]{Elmegreen1998}
  Elmegreen, B.G. 1998, in \asp, 148, Origins, ed. C.E. Woodward et al. (San Francisco: ASP), 150

\bibitem[Fukui et al.(2018)]{Fukui2018}
  Fukui, Y., et al. 2018, \pasj, 70, S46

\bibitem[Gennaro et al.(2012)]{Gennaro2012}
  Gennaro, M., et al. 2012, \aap, 542, A74
\bibitem[Guesten and Mezger(1982)]{Guesten1982}
  Guesten, R., \& Mezger, P.G. 1982, Vistas Astron., 26, 159
  
\bibitem[Habe and Ohta(1992)]{Habe&Ohta1992}
  Habe, A., \& Ohta, K. 1992, \pasj, 44, 203
\bibitem[Han et al.(2006)]{Han2006}
  Han, J.L., Manchester, R.N., Lyne, A.G., Qiao, G.J., \& van Straten, W. 2006, \apj, 642, 868
\bibitem[Hattori et al.(2016)]{Hattori2016}
  Hattori, Y., et al. 2016, \pasj, 68, 37
\bibitem[Heiles and Crutcher(2005)]{Heiles2005}
  Heiles, C., \& Crutcher, R. 2005, Cosmic Magnetic Fields (Lect. Notes Phys. vol.664), 137
\bibitem[Hou and Gao(2014)]{Hou&Gao2014}
  Hou, L.G., \& Gao, X.Y. 2014, \mnras, 438, 426
\bibitem[Hou and Han(2014)]{Hou2014}
  Hou, L.G., \& Han, J.L. 2014, \aap, 569, A125

\bibitem[Ishihara et al.(2011)]{Ishihara2011}
  Ishihara, D., Kaneda, H., Onaka, T., Ita, Y., Matsuura, M., \& Matsunaga, N. 2011, \aap, 534, A79
  
\bibitem[Kaneda et al.(2013)]{Kaneda2013}
  Kaneda, H., et al. 2013, \aap, 556, A92
\bibitem[Kawada et al.(2007)]{Kawada2007}
  Kawada, M., et al. 2007, \pasj, 59, S389
\bibitem[Kennicutt and Evans et al.(2012)]{Kennicutt2012}  
  Kennicutt, R.C., \& Evans, N.J. 2012, \araa, 50, 531
\bibitem[Kraemer et al.(2010)]{Kraemer2010}  
  Kraemer, K.E., et al. 2010, AJ, 139, 2319
\bibitem[Kokusho et al.(2017)]{Kokusho2017}
  Kokusho, T., Kaneda, H., Bureau, M., Suzuki, T., Murata, K., Kondo, A., \& Yamagishi, M. 2017, \aap, 605, A74
\bibitem[Kuchar and Bania(1994)]{Kuchar1994}
  Kuchar, T.A. \& Bania, T.M. 1994, \apj, 436, 117

\bibitem[Latter(1991)]{Latter1991}
  Latter, W.B. 1991, \apj, 377, 187
\bibitem[Lorimer et al.(2006)]{Lorimer2006}
  Lorimer, D.R., et al. 2006, \mnras, 372, 777
  
\bibitem[Mallick et al.(2013)]{Mallick2013}
  Mallick, K.K., Kumar, M.S.N., Ojha, D.K., Bachiller, R., Samal, M.R., \& Pirogov, L. 2013, \apj, 779, 113
\bibitem[Martins, Schaerer and Hillier(2005)]{Martins2005}
  Martins, F., Schaerer, D., \& Hillier, D.J. 2005, \aap, 436, 1049
\bibitem[Mathis, Mezger and Panagia(1983)]{Mathis1983}
  Mathis, J.S., Metzger, P.G., \& Panagia, N. 1983, \aap, 128, 212
\bibitem[Matsuura et al.(2009)]{Matsuura2009}
  Matsuura, M., et al. 2009, \mnras, 396, 918
\bibitem[Misiriotis et al.(2006)]{Misiriotis2006}
  Misiriotis, A., Xilouris, E.M., Papamastorakis, J., Boumis, P., \& Goudis, C.D. 2006, \aap, 459, 113
\bibitem[Miszalski et al.(2008)]{Miszalski2008}
  Miszalski, B., Parker, Q.A., Acker, A., Birkby, J.L., Frew, D.J., \& Kovacevic, A. 2008, \mnras, 384, 525
\bibitem[Molinari et al.(2010)]{Molinari2010}
  Molinari, S., et al. 2010, \pasp, 122, 314
\bibitem[Molinari et al.(2016)]{Molinari2016}
  Molinari, S., et al. 2016, \aap, 591, A149

\bibitem[Nakanishi and Sofue(2006)]{Nakanishi&Sofue2006}
  Nakanishi, H., \& Sofue, Y. 2006, \pasj, 58, 847
\bibitem[Nakanishi and Sofue(2016)]{Nakanishi2016}
  Nakanishi, H., \& Sofue, Y. 2016, \pasj, 68, 5 

\bibitem[Ohama et al.(2018)]{Ohama2018}
  Ohama, A., et al. 2018, \pasj, 70, S45
\bibitem[Onaka et al.(1996)]{Onaka1996}
  Onaka, T., Yamamura, I., Tanabe, T., Roellig, T.L., \& Yuen, L. 1996, \pasj, 48, L59
\bibitem[Onaka et al.(2007)]{Onaka2007}
  Onaka, T., et al. 2007, \pasj, 59, S401
\bibitem[Osterbrock (1989)]{Osterbrock1989}
  Osterbrock, D.E. 1989, Astrophysics of gaseous nebulae and active galactic nuclei, Mill Valley, CA: University Science Books

\bibitem[Parker et al.(2006)]{Parker2006}
  Parker, Q.A., et al. 2006, \mnras, 373, 79
\bibitem[Pavel and Clemens(2012)]{Pavel2012}
  Pavel, M.D., \& Clemens, D.P. 2012, \apj, 760, 150

\bibitem[Rahman and Murray(2010)]{Rahman2010}  
  Rahman, M., \& Murray, N. 2010, \apj, 719, 1104
\bibitem[Rodr{\'{\i}}guez-Esnard, Trinidad and Migenes (2012)]{Rodriguez2012}  
  Rodr{\'{\i}}guez-Esnard, T., Trinidad, M.A., \& Migenes, V. 2012, \apj, 761, 158
  
\bibitem[Samal et al.(2018)]{Samal2018}
  Samal, M.R., Deharveng, L., Zavagno, A., Anderson, L.D., Molinari, S., \& Russeil, D. 2018, A\&A, 617, A67
\bibitem[Simpson et al.(2012)]{Simpson2012}
  Simpson, R.J., et al. 2012, \mnras, 424, 2442
\bibitem[Smith et al.(2007)]{Smith2007}
  Smith, J.D.T., et al. 2007, \apj, 656, 770
\bibitem[Stierwalt et al.(2014)]{Stierwalt2014}
  Stierwalt, S., et al. 2014, \apj, 790, 124
\bibitem[Str$\ddot{\rm{o}}$mgren (1939)]{Stromgren1939}
    Str$\ddot{\rm{o}}$mgren, B. 1939, \apj, 89, 526


\bibitem[Takita et al.(2015)]{Takita2015}
  Takita, S., et al. 2015, \pasj, 67, 51
\bibitem[Tenorio-Tagle (1979)]{Tenorio1979}
  Tenorio-Tagle, G. 1979, \aap, 71, 59
\bibitem[Tielens(2008)]{Tielens2008}
  Tielens, A.G.G.M. 2008, \araa, 46, 289
\bibitem[Torii et al.(2015)]{Torii2015}
  Torii, K., et al. 2015, \apj, 806, 7

\bibitem[Watson et al.(2009)]{Watson2009}
  Watson, C., Corn, T., Churchwell, E.B., Babler, B.L., Povich, M.S., Meade, M.R., \& Whitney, B.A. 2009, \apj, 694, 546
\bibitem[Watson, Hanspal and Mengistu(2010)]{Watson2010}
  Watson, C., Hanspal, U., \& Mengistu, A. 2010, \apj, 716, 1478
\bibitem[Webber and Yushak(1983)]{Webber1983}
  Webber, W.R., \& Yushak, S.M. 1983, \apj, 275, 391
\bibitem[Whitworth et al.(2018)]{Whitworth2018}
  Whitworth, A., Lomax, O., Balfour, S., M{\`e}ge, P., Zavagno, A., \& Deharveng, L. 2018, \pasj, 70, S55
\bibitem[Wolfire et al.(2003)]{Wolfire2003}
  Wolfire, M.G., Mckee, C.F., Hollenbach, D., \& Tielens, A.G.G.M. 2003, \apj, 587, 278    

\bibitem[Zhang, Wang and Xu(2013)]{Zhang2013}
  Zhang, C.-P., Wang, J.-J., \& Xu, J.-L. 2013, \aap, 550, A117
\bibitem[Zinnecker and Yorke(2007)]{Zinnecker&Yorke2007}
  Zinnecker, H., \& Yorke, H.W. 2007, \araa, 45, 481

  
  
\end{thebibliography}
\end{document}